\pdfoutput=1
\newcommand*{\ATLASLATEXPATH}{latex/}
\documentclass[cernpreprint,txfonts,UKenglish,texlive=2016]{\ATLASLATEXPATH atlasdoc}
\usepackage[subfigure=true]{\ATLASLATEXPATH atlaspackage}
\usepackage{\ATLASLATEXPATH atlasbiblatex}

\makeatletter
\def\blx@maxline{77}
\makeatother

\usepackage{\ATLASLATEXPATH atlascontribute}

\usepackage[process=true]{\ATLASLATEXPATH atlasphysics}

\graphicspath{{logos/}{figures/}}

\usepackage{JetMetRatioXsec-defs}
\usepackage{rotating}
\usepackage{multirow}
\usepackage{pdflscape}

\usepackage{pifont}

\AtlasTitle{Measurement of detector-corrected observables sensitive to
the anomalous production of events with jets and large missing transverse momentum
in $\boldsymbol{pp}$ collisions at $\mathbf{\sqrt{s}=13}$~TeV using the ATLAS detector}

\usepackage{authblk}
\renewcommand\Authands{,\xspace} 
\renewcommand\Affilfont{\itshape\small} 
 \usepackage{authblk}
 \renewcommand\Authands{, } 
 \renewcommand\Affilfont{\itshape\small} 

\AtlasRefCode{EXOT-2016-03}

\PreprintIdNumber{CERN-EP-2017-116}

\AtlasJournal{EPJC}
\AtlasAbstract{
Observables sensitive to the anomalous production of events containing hadronic jets and missing momentum 
in the plane transverse to the proton beams at the Large Hadron Collider are presented. The observables are 
defined as a ratio of cross sections, for events containing jets and large missing transverse momentum to 
events containing jets and a pair of charged leptons from the decay of a $Z/\gamma^\ast$ boson. This definition 
minimises experimental and theoretical systematic uncertainties in the measurements. This ratio is measured differentially 
with respect to a number of kinematic properties of the hadronic system in two phase-space regions; one inclusive single-jet 
region and one region sensitive to vector-boson-fusion topologies. The data are found to be in agreement with the Standard Model 
predictions and used to constrain a variety of theoretical models for dark-matter production, including simplified models, effective 
field theory models, and invisible decays of the Higgs boson. The measurements use 3.2\,\ifb{} of proton--proton collision data 
recorded by the ATLAS experiment at a centre-of-mass energy of 13\,\TeV\ and are fully corrected for detector effects, 
meaning that the data can be used to constrain new-physics models beyond those shown in this paper.
}

\hypersetup{pdftitle={ATLAS document},pdfauthor={The ATLAS Collaboration}}

\begin{document}

\maketitle

\newcommand{\comment}[1]{\ding{110}\ding{43}\textcolor{blue}{Comment: #1}}
\newcommand{\todo}[1]{\ding{110}\ding{43}\textcolor{red}{TODO: #1}}
 \newcommand{\ifprefchar}{\ifpunctmark{'}}

\section{Introduction}
\label{sec:introduction}
The Standard Model of particle physics (SM) is an extremely successful
theory, describing the fundamental building blocks of nature and the
interactions between them. Despite
its many successes, it is
known that the SM does not provide a complete description: for example it does not
explain the abundance of dark matter in our universe, known to exist from astrophysical observations~\cite{Bertone:2004pz,deSwart:2017heh,Conrad:2017pms}.
One of the main aims of the physics programme at the Large Hadron Collider
(LHC)~\cite{Evans:2008zzb}  is to find evidence of new phenomena,
either via directly searching for the
signatures predicted by specific scenarios beyond the Standard Model (BSM) or, as is the case in this paper,
by performing a more general search for deviations from SM predictions.

\begin{figure*}
	\begin{center}
	\subfigure[]{
          \includegraphics[width=0.4\textwidth]{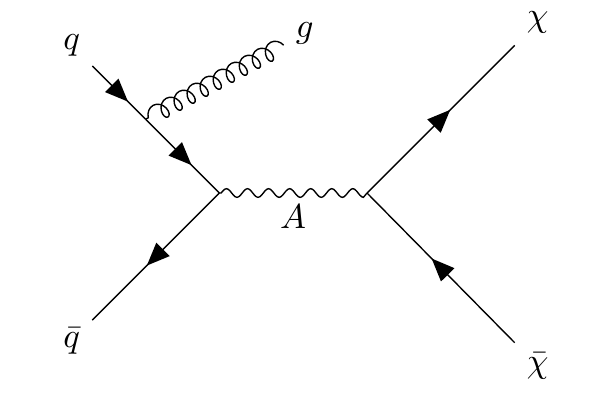}
          \label{fig:feynman-dm1}
	}
	\subfigure[]{
          \includegraphics[width=0.4\textwidth]{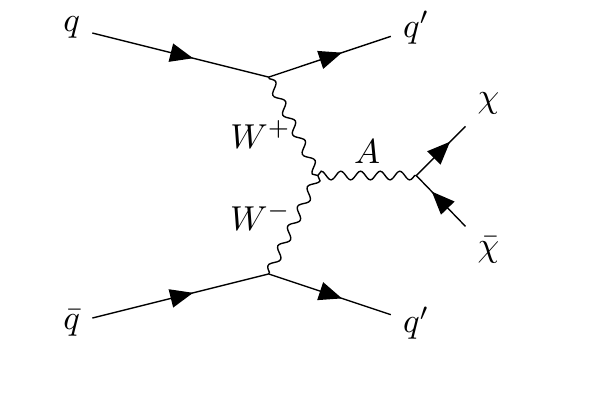}
          \label{fig:feynman-dm2}
	}
\subfigure[]{
          \includegraphics[width=0.4\textwidth]{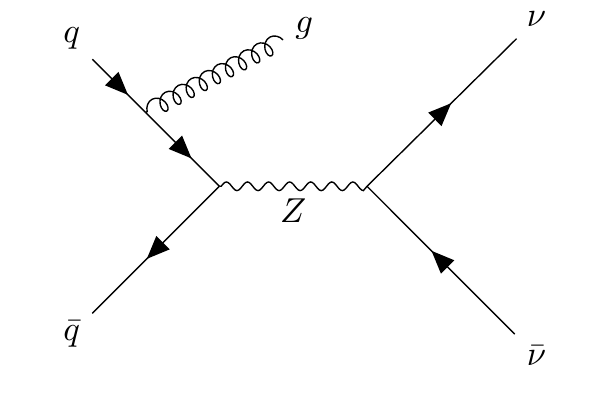}
          \label{fig:feynman-mono}
	}
\subfigure[]{
          \includegraphics[width=0.4\textwidth]{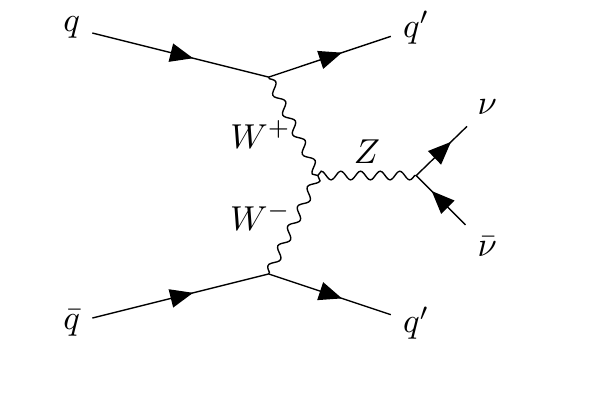}
          \label{fig:feynman-vbf}
	}
	\end{center}
	\caption {Example Feynman diagrams for WIMP $\chi$ pair production
          with mediator $A$ produced (a) in
          association with one jet and (b)  via vector-boson
          fusion. Example Feynman diagrams for the Standard Model
          background to (c) the process with one jet and (d) the vector-boson fusion process.} 
  \label{fig:feynmanDM}
\end{figure*}

New physics phenomena at the LHC may manifest themselves as events with
jets of collimated, mostly hadronic, particles and a momentum
imbalance in the plane transverse to the LHC beams, known as missing
transverse momentum, \ptmiss. The \ptmiss{} may indicate the
presence of particles that do not interact via the strong or
electromagnetic interactions and therefore cannot be directly detected
in the LHC detectors. These particles are referred to as invisible. 
In particular, new-physics models 
predicting the existence of weakly interacting massive particles (WIMPs), dark-matter candidates that could be produced at the LHC,
could lead to such a signature~\cite{Abdallah:2015ter}. As an example, a
Feynman diagram is shown in Figure~\ref{fig:feynman-dm1}, where a
mediator, $A$, is produced in association with a gluon-initiated jet and decays to a WIMP pair ($\chi\bar{\chi}$).
Limits have previously been placed in such models by comparing the number of events in \ptmissjet{}
final states in LHC data with the number of background events expected to be seen in the 
detector (the detector level)~\cite{EXOT-2015-03,CMS-EXO-12-048}.
Another possible production mechanism for the experimental
observation of weakly interacting BSM particles is vector-boson fusion
(VBF)~\cite{Cotta:2012nj}, as shown in
Figure~\ref{fig:feynman-dm2}. This is a topology similar to that
in the invisible decay of a VBF-produced Higgs
boson~\cite{Shrock:1982kd,Choudhury:1993hv,Eboli:2000ze}, for which
limits have previously been
set~\cite{HIGG-2013-16,Khachatryan:2016whc} using detector-level data.
The dominant SM process leading to the same final states is
the production of a \Z{} boson in association with jets,
where the \Z{} boson decays to a pair of neutrinos. Example diagrams
are shown in Figures~\ref{fig:feynman-mono} and~\ref{fig:feynman-vbf}.

This paper presents a measurement of differential observables that are
sensitive to the anomalous production of events containing one or more
hadronic jets with high transverse momentum, \pt,  produced in
association with a large \ptmiss{}. The measurements are performed using data corresponding to an
integrated luminosity of 3.2~\ifb{} of proton--proton
collisions at $\sqrt{s}=13$\,\TeV, collected by the ATLAS
detector~\cite{PERF-2007-01} in 2015.
The observables  are corrected for detector inefficiencies and
resolutions and are presented at the
particle level. 
They are constructed from a ratio of cross-sections,
\begin{equation*}
\Rmiss = \frac{\sigma_{\textrm{fid}} \left( \ptmissjet{} \right)} {\sigma_{\textrm{fid}}  \left(\lljet{} \right)},
\end{equation*}
defined in a fiducial phase space. 
The numerator is the fiducial cross-section for \ptmissjet{} events, which corresponds to the fiducial cross-section 
for inclusive $Z(\to\nu\bar{\nu})+$ jets production in the SM.
The denominator is the fiducial cross-section for \lljet{} events, where the unobserved 
system that produces the \ptmiss{} in the numerator is replaced by an observed,
opposite-sign, same-flavour pair of charged leptons consistent with originating from a $Z/\gamma^\ast$ boson.
The lepton pair can be either a pair of electrons or muons. 
The jet system is required to satisfy very similar selection criteria in both the 
\ptmissjet{} and \lljet{} samples of events so as to significantly reduce experimental 
and theoretical uncertainties in the ratio measurement.
The presence of BSM physics in the numerator would lead to a discrepancy between the 
measured ratio and that predicted by the SM.

The approach used in this paper allows for direct comparison of 
SM and BSM predictions at the particle level, without the need to simulate the effects of the ATLAS detector.
This is computationally efficient and enables those without access to a precise simulation of 
the ATLAS detector to compare the data with
predictions from alternative BSM models as they become available.
Since each alternative BSM model may predict event signatures with
different kinematic properties, the publication of the kinematic
distributions enhances the usefulness and longevity of the data.
Furthermore, future improvements in the predictions of the SM processes that
contribute to the ratio can be compared to the particle-level data and
limits in BSM models can be updated accordingly.

Particle-level measurements of SM processes are common in collider
physics and have, on occasion, been used to set limits in BSM
models (see e.g.~\cite{STDM-2012-03}), although not to search for new physics in the \ptmissjet{} final state. 
Moreover, a measurement of the particle-level ratio allows the denominator to provide a constraint on
the dominant SM process contributing to the \ptmissjet{} final state.  
Many sources of systematic uncertainty cancel in the ratio because the
requirements on the hadronic system and the definition of  the measured kinematic
variables, determined from the hadronic system, are similar in the
numerator \ptmissjet{} and denominator \lljet{} events. 
This is made possible by treating the identified charged leptons in
\lljet{} events as invisible when calculating the \ptmiss{}.
This cancellation occurs, for example, for phenomenological
uncertainties in the prediction of initial-state parton radiation and
 experimental uncertainties in the jet reconstruction, energy scale and resolution.

The ratio measurements are presented in two phase-space regions: the
\onejet{} region, containing  at least one high-\pt{} jet, and the \vbf{}
region, containing at least two  high-\pt{} jets, and satisfying additional
selection criteria to enhance the \vbf{} process.
This ratio is measured as a function of a number of kinematic properties of the hadronic system of the event 
and the statistical and systematic correlations between the different distributions are determined. 
The data and correlation information are made publicly available.

The remainder of this paper is laid out as follows.
The ATLAS detector and event reconstruction are described in
Section~\ref{sec:methods-detector}.
The fiducial regions defined by particle-level objects and event
selections, together with the measured variables,
are detailed in Section~\ref{sec:particle}. The \ptmissjet{} and
\lljet{} event samples are
selected as described in Section~\ref{sec:reco}. 
Samples of events were produced with Monte Carlo event generators and are used to correct the data for detector
effects, to estimate background and signal contributions, and to assign
systematic uncertainties to the results. Details of these samples are given in Section~\ref{sec:MC}.
Predicted
backgrounds, explained in Section~\ref{sec:backgrounds}, are
subtracted from the selected data and the ratio is computed. A~correction for
detector effects is applied to the ratios, as described in
Section~\ref{sec:corrections}, so that they are defined at 
particle level with the definitions from Section~\ref{sec:particle}.
Systematic
uncertainties in the measurement and theoretical predictions are
summarised in Section~\ref{sec:syst}. The detector-corrected events 
in the electron and muon channels are
combined to form particle-level ratios to \lljet{} events, as
described in Section~\ref{sec:combo}. These are
compared to the expected SM ratios and to the expected ratios including
example BSM models in Section~\ref{sec:results}. 
The results are discussed in Section~\ref{sec:discuss} and example limits are placed on BSM model parameters.
Finally, conclusions are given in Section~\ref{sec:conclusions}.

\section{ATLAS detector and event reconstruction}
\label{sec:methods-detector}
The ATLAS detector~\cite{PERF-2007-01,ATLAS-TDR-19} is a multipurpose particle detector with a
cylindrical geometry. 
ATLAS consists of
layers of tracking detectors, calorimeters, and muon chambers. 
The inner detector (ID) 
covers the
pseudorapidity range $|\eta| < 2.5$.\footnote{ATLAS uses a right-handed coordinate
  system with its origin at the nominal interaction point in the
  centre of the detector and the $z$-axis along the beam pipe. The
  $x$-axis points to the centre of the LHC ring, and the $y$-axis
  points upward. Cylindrical coordinates $(r,\phi)$ are used in the
  transverse plane, $\phi$ being the azimuthal angle around the beam
  pipe. The pseudorapidity is defined in terms of the polar angle
  $\theta$ as $\eta=-\ln[\tan(\theta/2)]$. Rapidity is defined as 
  $y=0.5 \ln[(E+p_z)/(E-p_z)]$ where $E$ denotes the energy and $p_z$ 
  is the momentum component along the beam direction.} 
The ID is immersed in a 2~T magnetic field and measures the
trajectories and momenta of charged particles.
The calorimeter covers the
pseudorapidity range $|\eta| < 4.9$.
 Within $|\eta| <2.47$, the finely segmented electromagnetic calorimeter identifies electromagnetic showers and measures
their energy and position, providing electron identification together with the ID. 
The muon spectrometer (MS) surrounds the calorimeters and provides muon identification and
measurement in the region $|\eta| < 2.7$.

Jets are reconstructed from energy deposits in the calorimeters, 
using the anti-$k_t$ jet algorithm~\cite{Cacciari:2008gp,Cacciari:2011ma},
with a jet-radius parameter of 0.4. The measured jet \pt{} is corrected~\cite{Aaboud:2017jcu}
for the detector response and contributions to the jet energy from multiple 
proton--proton interactions (pileup). 
Jet quality selection criteria~\cite{ATLAS-CONF-2015-029} are applied.
Track-based variables are then used to suppress
jets with $|\eta|<2.4$ and $\pt < 50$\,\GeV\ 
by requiring that a significant fraction of the tracks associated with each jet 
must have an origin compatible with the primary vertex in the event, which further suppresses jets from pileup interactions.

A muon is reconstructed by matching a track (or track segment)
reconstructed in the MS to
a track reconstructed in the ID.  
Its momentum is calculated
by combining the information from the two systems and correcting for energy deposited in the
calorimeters. 
Quality requirements are applied using the \emph{loose} working point 
as described in Ref.~\cite{PERF-2015-10}.
An electron is reconstructed from an energy deposit (cluster) in the
electromagnetic calorimeter matched to a track in the ID. Its momentum
is computed from the cluster  energy and
the direction of the track.
Electrons are distinguished from other particles using several
identification criteria that rely on the shapes of electromagnetic
showers as well as tracking and track-to-cluster matching
quantities. The output of a likelihood function taking these
quantities as input, similar to that
described in  Ref.~\cite{PERF-2016-01}, and using the \emph{loose}
working point described therein, is used to identify
electrons. 
Data-driven energy/momentum scale corrections~\cite{PERF-2015-10} are applied to both reconstructed muons and electrons.
Leptons are required to be
associated with the primary vertex, defined as the vertex with the
highest $\Sigma\pt^2$ of its associated tracks, in order to suppress leptons originating from pileup and secondary decays.
Hadronic decays of $\tau$ leptons ($\tau \to$ hadrons $+ \nu$) are predominantly characterised by the presence 
of one or three charged particles and possibly neutral pions.
A multivariate boosted decision tree  identification, based on
calorimetric shower shape and track multiplicity of the $\tau$ candidates, 
is used to reject jets faking $\tau$ leptons. More details are
given in Ref.~\cite{ATL-PHYS-PUB-2015-045}, with the \emph{loose} working point being
used in this analysis.

The \ptmiss{} is reconstructed as the magnitude of the negative vector
sum of the transverse momenta of all detected particles, as described in Ref.~\cite{PERF-2014-04}. 
The \ptmiss{} calculation uses a soft term that is calculated using tracks
within the ID which are not associated with jets or with leptons that
are being treated as invisible particles.
The momenta of calibrated jets with 
$\pt > 20$\,\GeV\ are used. 

Events in the numerator and the \mumu{} denominator are selected by
a trigger that requires $\ptmiss > 70$\,\GeV, as computed in
the final stage of the two-level trigger system. 
Since the momenta from muons are not included in the \ptmiss{}
calculation in this trigger, the muons appear to the trigger as invisible particles 
and hence the trigger can also be used to select \mumu{} events.
This trigger is 100\,\% efficient for the offline $\ptmiss > 200$\,\GeV\ requirement used in the analysis. Events in
the \ee{} denominator are selected by a single-electron trigger, 
with an efficiency ranging between 93\,\% and more than 99\,\% for 
electrons with $\pt>80$\,\GeV, depending on their pseudorapidity.

\section{Particle-level objects, event selections and measured variables}
\label{sec:particle}
The detector-corrected data are presented in fiducial regions defined in this
section. The definition of the measured variables is also given.
The final state of an event is defined using all particles with $c\tau$ longer than 10\,mm. 
Final-state particles that interact via the
strong or electromagnetic interactions are referred to as
visible particles, whereas those that interact via neither are referred 
to as invisible particles.

At particle level, the \lljet{} events for the denominator of \Rmiss{} are required to
have exactly one opposite-sign, same-flavour pair of prompt\footnote{Prompt refers to particles not coming from the
decay of a hadron or from the decay of a $\tau$ lepton.} leptons: an $e^+e^-$ or
$\mu^+\mu^-$ pair. The four-momenta of prompt photons
within a cone of $\Delta R= \sqrt{(\Delta\eta)^2 + (\Delta\phi)^2} = 0.1$
around each lepton
are added to the four-momenta of the leptons and then removed from
the final state, as motivated in Ref.~\cite{ATL-PHYS-PUB-2015-013}.
These so-called `dressed' leptons are required to satisfy the kinematic
criteria detailed below. 

Both the numerator and denominator of \Rmiss{} are
required to satisfy a number of phase-space-dependent criteria,
summarised in Table~\ref{tab:fiducial}.
The fiducial phase-space definitions are motivated by the
acceptance of the detector and the
trigger~\cite{TRIG-2016-01}, background reduction and, in the case of the \vbf{} phase space,
by the enhancement of the contribution from VBF processes. 
The $\ptmiss$ value is defined as the magnitude of the negative vector sum of the transverse momenta of all visible 
final-state particles with $|\eta|<4.9$, as this corresponds to the
edge of the calorimeter. Muons with $|\eta| > 2.5$ are
excluded as they
contribute only negligibly to the calculation of \ptmiss{} in this analysis, via a small
energy deposition in the calorimeter.
For the denominator, the $\ptmiss$ variable is modified: the selected
dressed leptons are excluded from the vector sum, making the variable very similar between numerator and denominator.
Jets are
reconstructed with the anti-$k_t$ jet algorithm with jet radius
parameter 0.4, excluding
invisible particles and muons.

The event-level veto on (additional) leptons is applied to reduce the contribution
from background processes. In particular, this requirement
significantly reduces the background to \ptmissjet{} events from \W{}
bosons produced in association with jets.
The requirement on
the difference in azimuthal angle between \ptmiss{} and any of the
leading four jets with $\pt > 30$\,\GeV,
\dphiptmissj{}, suppresses backgrounds from multijet events, as is
discussed in Section~\ref{sec:backgrounds}.
For the denominator, the minimum \pt{} requirement for the leading lepton is much larger than the
subleading lepton as events with a large \ptmiss{} tend to have
one very high \pt{} lepton. The subleading lepton \pt{} can be much lower,
in particular if it is in the direction opposite the
decaying \Z{} boson. 
The leading lepton $\pt$ tends to be lower in
$t\bar{t}$ events, motivating the choice to make an asymmetric requirement.
The requirement on the dilepton invariant mass to be between
66 and 116\,\GeV\ is implemented to minimise the contribution of the photon
propagator and interference terms in the
denominator, making it as similar as possible to the numerator.

In VBF, at least two jets
are in the final state and, due to the colourless exchange, less hadronic
activity in the rapidity space between the two jets is expected,
which motivates the central-jet veto. The
 dijet invariant mass (\mjj{}) requirement suppresses the contribution
from diboson events where one boson decays hadronically. 

\begin{table}
\centering
\caption{\label{tab:fiducial} Definitions for the
\onejet{} and \vbf{} fiducial phase spaces. Here \mjj{}
is the invariant mass of the two leading (in \pt{}) jets, \dphiptmissj{} is the difference in azimuthal
angle between \ptmiss{} and a jet axis. The lepton veto is applied to
events in the numerator (denominator) of \Rmiss{} containing at least one (three)
prompt lepton(s) or lepton(s) from $\tau$ decays.
The selected leptons in the denominator are treated as invisible when calculating the \ptmiss{} value.
The central-jet veto is applied to
any jets in the rapidity ($y$) space between the two leading jets. The
dilepton invariant mass is denoted by \mll{}. }
\begin{tabular}{ l|c|c } 
 \hline
 \textbf{Numerator and denominator}                           & \onejet{}          & \vbf{} \\ 
\hline
\ptmiss{}                  & \multicolumn{2}{c}{$> 200$\,\GeV }\\
(Additional) lepton veto   & \multicolumn{2}{c}{ No $e,\mu$ with $\pt > 7$\,\GeV, $|\eta| < 2.5$} \\ 
Jet $|y|$                  & \multicolumn{2}{c}{$< 4.4$}        \\
Jet \pt{}                  & \multicolumn{2}{c}{$> 25$\,\GeV}        \\
\dphiptmissj{}             & \multicolumn{2}{c}{$> 0.4$, for the four leading jets with $\pt > 30$\,\GeV}\\
\hline
Leading jet \pt{}          & $> 120$~GeV & $> 80$\,\GeV \\
Subleading jet \pt{}      &  --                 & $> 50$\,\GeV \\
Leading jet $|\eta|$       &  $< 2.4$         &  --        \\
\mjj{}                     & --                  &  $> 200$\,\GeV \\
Central-jet veto           & --                  & No jets with $\pt > 25$\,\GeV \\
 \hline
 \textbf{Denominator only}                           &
\multicolumn{2}{c}{\onejet{} and \vbf{}}\\ 
\hline
Leading lepton \pt{}          &  \multicolumn{2}{c}{ $> 80$\,\GeV }    \\
Subleading lepton \pt{}   &  \multicolumn{2}{c}{ $> 7$\,\GeV }    \\
Lepton $|\eta|$   &  \multicolumn{2}{c}{ $< 2.5$ }    \\
\mll{}                  & \multicolumn{2}{c}{ 66--116\,\GeV }    \\
$\Delta R$ (jet, lepton) & \multicolumn{2}{c}{$> 0.5$, otherwise jet is removed}\\
\hline
\end{tabular}
\end{table}

In order to increase the sensitivity to a range of targeted BSM
scenarios, four differential measurements of \Rmiss{} are made with respect to: 
\ptmiss{} in the \onejet{}  and \vbf{} phase spaces, as well as \mjj{} and \dphijj{}
in the \vbf{} phase space, 
where \dphijj{} is the difference in
azimuthal angle between the two leading jets.
Due to the larger mediator mass and higher energy scale of the interaction, many BSM signatures tend
to have harder \ptmiss{} distributions than the SM processes, meaning
that sensitivity to these models is enhanced in the high-\ptmiss{} region. 
Since the \vbf{} process leads to events with
a harder \mjj{} spectrum than processes involving the strong production
of dijets, the high-\mjj{} region gives more discriminating power for
VBF models. The expected \dphijj{} distribution  varies between different
BSM theories and could therefore give additional sensitivity and
possibly help to distinguish between models, should a signal be seen.

\section{Detector-level event selection}
\label{sec:reco}
Events are required to contain a primary vertex
with at least two associated tracks, each with $\pt > 400$\,\MeV.
Events containing a jet with $\pt > 20$\,\GeV\ not originating from a proton--proton
interaction are rejected. Such jets are identified by jet quality selection criteria involving quantities such as the
pulse shape of the energy depositions in the cells of the calorimeters, electromagnetic fraction in
the calorimeter, calorimeter sampling fraction, or the fraction of
energy coming from charged particles.

The kinematic selection criteria given in Table~\ref{tab:fiducial}
are identically applied to detector-level objects, with an additional
exclusion of electrons in the region $1.37 < |\eta| < 1.52$, which corresponds to the
calorimeter barrel--endcap transition region, and in the region  $2.47
< |\eta| < 2.5$, since electrons are identified only for $|\eta| <2.47$.
All electrons, as well as muons used for the lepton veto, are required to be
isolated from other particles. In both cases, the LooseTrackOnly
isolation working points described in Refs.~\cite{PERF-2015-10,PERF-2016-01} are used.
A veto on events
containing an identified hadronically decaying $\tau$ lepton, with the
total $\pt$ of the visible decay products being greater than 20\,\GeV,
is also applied to reduce the contribution from \Wtaunu{}
events to \ptmissjet{} events. This veto is not applied at the
particle level due to the complication of defining a hadronically
decaying $\tau$ lepton in terms of stable final-state particles.

In this analysis, identified charged leptons
are either vetoed or treated as invisible particles in the \ptmiss{}
calculation. In particular, for the \lljet{} denominator, the measured momenta of selected
electrons, muons, and jets close to muons which are consistent with
being associated with final-state radiation photons clustered close to the
muon ID track, are treated as invisible. A jet is considered to be consistent with a
final-state photon if its transverse momentum is less than twice the
transverse momentum of the associated muon and it has fewer than five
associated ID tracks.
This makes \ptmiss{} very similar between numerator and denominator.

\section{Monte Carlo simulation}
\label{sec:MC}
Events containing \Z{} and \W{} bosons (collectively termed $V$) were generated using Monte Carlo (MC) event generators.
Samples contributing to inclusive $Z+$jets production (\Znunu{}, \Zgll{} and diboson $ZV$, where the \Z{} decays to
a $\nu\bar{\nu}$, $e^+e^-$ or $\mu^+\mu^-$ pair and $V$ is a hadronically decaying \W{} or \Z{} boson)
are used for the detector corrections. 
Samples of \Wlnu{} (including $WV$ where the \W{} decays leptonically
and the $V$ decays hadronically),
top--antitop quark pairs, single-top-quark and leptonically decaying diboson ($WW$, $WZ$, $ZZ$) events are
used to estimate backgrounds.

Events containing single \Z{} and \W{} bosons in association with jets were simulated using the \sherpa{} v2.2.0
event generator~\cite{Gleisberg:2008ta}. Matrix elements were calculated for up to two additional parton emissions 
at next-to-leading-order (NLO) accuracy and up to four additional parton emissions at leading-order (LO) accuracy
using the Comix~\cite{Gleisberg:2008fv} and OpenLoops~\cite{Cascioli:2011va} matrix element generators and merged 
with the \sherpa{} parton shower~\cite{Schumann:2007mg}, which is based on Catani--Seymour subtraction terms.
The merging of multi-parton matrix elements with the parton shower is achieved using an improved 
CKKW matching procedure~\cite{Catani:2001cc,Hoeche:2009rj}, which is extended to NLO accuracy using the 
MEPS@NLO prescription~\cite{Hoeche:2012yf}.
The \textsc{NNPDF3.0}nnlo parton distribution function (PDF) set~\cite{Ball:2014uwa} was used in conjunction with 
the dedicated parton-shower tuning developed by the \sherpa{} authors.
These $V+$jets samples were produced with a simplified scale-setting prescription in the multi-parton matrix elements
to improve the event generation speed. A theory-based reweighting of the jet-multiplicity distribution is applied, 
derived from event generation with the strict scale prescription.
The samples are normalised to a next-to-next-to-leading-order (NNLO)
prediction~\cite{Anastasiou:2003ds}. The full set-up is described in
detail in Ref.~\cite{ATL-PHYS-PUB-2016-003}.
Electroweakly produced $V+$jets as well as diboson production were generated 
using \sherpa{} v2.1.1 in conjunction with 
the CT10nlo~\cite{Lai:2010vv} PDF set and the dedicated parton-shower tuning developed by the \sherpa{} authors. 
The full set-up is described in detail in Ref.~\cite{ATL-PHYS-PUB-2016-002}.

Alternative samples of events with $V+$jets simulated using \mgfive{} v2.2.2~\cite{Alwall:2011uj} 
at LO and interfaced to the \pythia{} v8.186~\cite{Sjostrand:2007gs} parton shower 
are used for cross-checks and for the determination of systematic uncertainties.
The ATLAS A14 set of tuned parameters~\cite{ATL-PHYS-PUB-2014-021} is used together with 
the NNPDF3.0nlo PDF set. 
These samples are also normalised to the NNLO prediction.

Top--antitop pair production~\cite{Alioli:2011as}, as well as single-top-quark production
in the $Wt$~\cite{Re:2010bp} and $s$-channels~\cite{Frederix:2012dh,Alioli:2009je}, were generated using the 
Powheg-Box v2~\cite{Frixione:2007vw,Nason:2004rx,Frixione:2007nw} event generator with the CT10nlo PDF set for the matrix element 
calculations. Single-top $t$-channel events were generated using the Powheg-Box v1 event generator. 
Parton showering, hadronisation, and the underlying event were provided by \pythia{} v6.428~\cite{Sjostrand:2006za} using the 
CTEQ6L1 PDF set~\cite{Pumplin:2002vw} and the Perugia 2012 (P2012) set of tuned parton-shower parameters~\cite{Skands:2010ak}. 
The full set-up of these top-quark samples is described in detail in Ref.~\cite{ATL-PHYS-PUB-2016-004}.
The top-pair samples are normalised to a calculation at NNLO accuracy including soft-gluon resummation at next-to-next-to-leading 
logarithmic (NNLL) accuracy~\cite{Czakon:2011xx}.
The single-top samples are normalised using an NLO calculation including the resummation of soft gluon terms at
NNLL accuracy~\cite{Kidonakis:2010ux,Kidonakis:2010tc,Kidonakis:2011wy}.

WIMP simplified signal models were simulated using Powheg-Box v2 (r3049) using the model described 
in Ref.~\cite{Haisch:2013ata}.
This model implements the production of WIMP pairs with $s$-channel spin-1 mediator exchange at NLO precision.
Events were generated with the NNPDF3.0nlo PDF set with parton showering using \pythia{} v8.205~\cite{Sjostrand:2014zea} with 
the A14\,\cite{ATL-PHYS-PUB-2014-021} parameter set.
This model has a coupling $g_q$ of the SM quarks to the mediator, and a coupling $g_{\chi}$ of dark-matter particles to the mediator.
Couplings were set to a constant value of $g_q = 0.25$ and $g_{\chi}=1$, as recommended in Ref.~\cite{Abercrombie:2015wmb}.
A grid of samples was produced
for WIMP masses ranging from 1\,\GeV\ to 1\,\TeV\ and axial-vector mediator masses
between 10\,\GeV\ and 2\,\TeV.
More details of the samples are given in Ref.~\cite{EXOT-2015-03}.

In order to assess the sensitivity to invisible decays of the Higgs boson, $H \to ZZ \to 4\nu$ events were 
simulated using Powheg-Box v1~\cite{Alioli:2008tz,Nason:2009ai,Bagnaschi:2011tu} with
CT10 PDFs, and \pythia{} v8.165 simulating the parton shower, hadronisation and underlying
event. The cross-sections and their uncertainties for Higgs boson
production via vector-boson fusion, gluon--gluon fusion, and associated production
are taken from Ref.~\cite{Heinemeyer:2013tqa}. 

In order to search for general signatures of Dirac-fermion dark-matter coupling to weak bosons,
an implementation~\cite{Brooke:2016vlw} of an effective field theory~\cite{Cotta:2012nj} (EFT) 
in \textsc{FeynRules} v2.3.1~\cite{Alloul:2013bka} was used,
with \textsc{MadGraph5} v2.2.3~\cite{Alwall:2011uj} used to simulate
the hard interaction. This EFT includes ten possible dimension-five to dimension-seven operators
with a range of possible Lorentz structures, including some with different charge-parity ($CP$) properties for the
effective interaction between weak bosons and a dark matter candidate.
This model was interfaced to \pythia{} v8.212 
with the A14 parameter set and the NNPDF23LO~\cite{Ball:2012cx} PDF to simulate the effects
of parton showering, hadronisation and the underlying event.

All SM MC simulation samples were passed through \textsc{GEANT4}~\cite{geant-1,geant-2} 
for a full simulation~\cite{SOFT-2010-01} of the detector and are then reconstructed 
using the same analysis chain as the data.
Scale factors are applied to the simulated events to correct for the
small differences from data in the trigger, reconstruction,
identification, isolation, and impact parameter efficiencies for
leptons~\cite{PERF-2015-10,PERF-2016-01}. Furthermore, the
lepton and jet momentum scales and
resolutions are adjusted to match the data.
Additional proton--proton collisions in the same bunch crossing
 are overlaid. These are based on soft strong-interaction processes simulated
with \pythia{} v8.186 using the MSTW2008lo 
PDF set~\cite{Martin:2009iq} along with the A2 set of tuned parton-shower
parameters~\cite{ATL-PHYS-PUB-2011-014}.
The average number of proton--proton interactions per bunch crossing in this data set is 13.7.

\section{Backgrounds}
\label{sec:backgrounds}

The dominant background in the \ptmissjet{} numerator is from events
containing a leptonically decaying \W{} boson produced in association with
jets, which contain \ptmiss{} associated with an invisible particle: in this case the neutrino in the \W{} decay. 
Such events would pass the veto on additional leptons if the charged lepton ($e, \mu$ or
$\tau$) is not reconstructed or is outside the acceptance of the detector.
This background
includes contributions where the \W{} boson originates from a top-quark decay or diboson
events. The top-quark decay contribution to the \W{} background amounts to approximately
18\,\% (14\,\%) in the \onejet{} (\vbf{}) phase spaces.
The three lepton decay channels of 
the \W{} background contribute approximately 18\,\% (\Wmunu), 12\,\% (\Wenu) 
and 15\,\% (\Wtaunu) to the numerator.
The size of the combined \W{} background is similar to the SM \Znunu{}
contribution to the numerator at low \ptmiss{}, becoming less
important at high \ptmiss{}.

The contribution from this background is estimated using two \W{} control regions. A \Wmunu{} (\Wenu) control region is
selected by requiring a muon (electron) that is isolated from other
particles, with $\pt{} > 25$\,\GeV. 
The requirements on the jets, \ptmiss{}, and the veto on additional
leptons are identical to
those of the \ptmissjet{} signal region.
In the \Wmunu{} control region, the muon is treated as an invisible particle in
the \ptmiss{} calculation, in order to make the region as similar as
possible to the signal region. This is because the signal region has a
veto on reconstructed muons and so the muon is often not included in
the \ptmiss{} calculation.
In the \Wenu{} control region, the energy of the electron is
included in the \ptmiss{} calculation, calibrated as a jet. This is
because the electron is usually
included in the signal region for \Wenu{} events, where the electron is
generally inside the acceptance of the calorimeter, but is not
identified, as a veto on identified electrons is applied in the
signal region.
\Wtaunu{} events, where the $\tau$ decay includes a muon (electron), are included in the
\Wmunu{} (\Wenu{}) control regions so that the contribution of these
events to the signal region is also included in this estimate.

The data in the \Wmunu{} and \Wenu control regions are collected using the \ptmiss\ 
and single-electron triggers discussed in Section \ref{sec:reco} and are corrected for
lepton inefficiencies on an event-by-event basis using \pt- and $\eta$-dependent lepton reconstruction, 
identification and isolation efficiencies, $\epsilon$, 
that were previously determined from data~\cite{PERF-2015-10,PERF-2016-01}. The data in 
the \Wenu control region are also corrected for the single-electron trigger inefficiency. 
A small background contribution from multijet events in the control region is estimated using 
dedicated MC simulation and subtracted from the data.
The efficiency- and multijet-corrected data are then used to predict the contribution from
\Wmunu{} and \Wenu{} events in the signal, which contains two types of events:
those for which the lepton is inside the
detector acceptance with $\pt{} > 7$\,\GeV\ but does not pass the lepton 
reconstruction and identification criteria,
and those with a lepton that is outside of the detector acceptance or
has $\pt{} < 7$\,\GeV. The in-acceptance contribution is determined for
each bin of a given distribution from the efficiency-corrected data in
the control region by applying an additional weight of 
$\left( 1 -  \epsilon\right)$ per event as well as correcting for the small
difference in lepton fiducial acceptance between the control region and
the signal region, using an acceptance-correction factor that is estimated using MC simulation.
The out-of-acceptance contribution is obtained by extrapolating
efficiency-corrected in-acceptance data using again acceptance corrections
derived from simulation.
As a cross-check, the \W{} background estimate is also determined using an alternative
method, described in Ref.~\cite{EXOT-2015-03}, where no
efficiency weights are applied to data and the simulation is used to extrapolate from 
the control region to the signal region. Compatible results are found.

There is no specific  \Wtaunu{} control region for hadronically
decaying taus, as it is difficult to
obtain a pure sample of  \Wtaunu{} events in data. Instead, background predictions for 
\Wtaunu{} with hadronically decaying $\tau$ leptons are obtained by reweighting the 
simulated \Wtaunu{} events, in each bin of each distribution, by the ratio of 
efficiency-corrected data to simulation determined in the \Wmunu{} or \Wenu control regions. 
The midpoint of the two predictions, obtained using the two control regions, is taken as the final 
\Wtaunu{} prediction and the difference between the midpoint and the
two predictions is taken as a systematic uncertainty. This choice is
made because a hadronically decaying $\tau$ lepton is often included in
the \ptmiss{} calculation, calibrated as a jet, which is similar to
the \Wenu{} control region. However, the $\tau$ decay includes a
neutrino, meaning that some part of it is invisible, which is similar
to the \Wmunu{} control region.

A much smaller background to the \ptmissjet{} events arises from multijet
events in which one or more jets are mismeasured leading to a
large measured \ptmiss. This implies that the \ptmiss{} direction is likely to point
towards one of the
jets and so most of this background is removed by
the \dphiptmissj{} requirement. The remaining
background is estimated  using a control region where at least
one of the four leading jets satisfies the criterion $\dphiptmissj < 0.1$.
A large multijet data sample is obtained from events selected with
single-jet triggers. These control events are required to be well
measured, meaning that the \ptmiss{} is low. In order to obtain a
sample of events that pass the \ptmiss{} selection, the jets in these
events are smeared 25\,000 times per multijet control event, according to the
full jet response distribution. This sample is used to extrapolate
between the control region and the signal region.
The multijet background amounts
to 2\%{} in the first \ptmiss{} bin, rapidly becoming negligible in
the higher \ptmiss{} bins. 
The small (0.5\%{}) \Zgll{} background to the  \ptmissjet{}
events is estimated using MC simulation. 

The background to \lljet{} events is dominated by top--antitop quark pairs,
with smaller contributions from diboson, single-top-quark, \W{} + jet and $\Z\to\tau^{+}\tau^{-}$
events. These backgrounds are all estimated with MC simulation
together with a control region that selects differently flavoured 
\lljet{} events (an $e^{\pm}\mu^{\mp}$ pair). All other selection criteria are the
same. This control region removes the contribution from
same-flavour \lljet{} events but retains contributions from the
background processes. Discrepancies between data and simulation of up to 50\,\% are
seen in the control region, depending on the phase space and the kinematic region.
A reweighting factor is found by fitting a polynomial to the ratio of
data to simulation in the control region and is applied to the
background contribution in the signal region. The full difference
between the background prediction with and without this reweighting is
taken as a systematic uncertainty.

Figures~\ref{fig:met-datamc} and~\ref{fig:mjj-dphi-datamc}  compare
detector-level data to 
MC simulation of \Znunu{} and \Zll{} events, plus estimated
backgrounds for selected \ptmissjet{} and selected \lljet{}
events in the signal region. Distributions of  \ptmiss{} in the \onejet{} and \vbf{}
phase spaces and for \mjj{} and \dphijj{} in the \vbf{} phase space
are compared.
For both the \ptmissjet{} and \lljet{} event rates, the data are above the
predictions from MC simulation and estimated
backgrounds. However, they are consistent within the 
systematic uncertainties, which are discussed in Section~\ref{sec:syst} in more detail.
\begin{figure*}
	\begin{center}
	\subfigure[]{
          \includegraphics[width=0.47\textwidth]{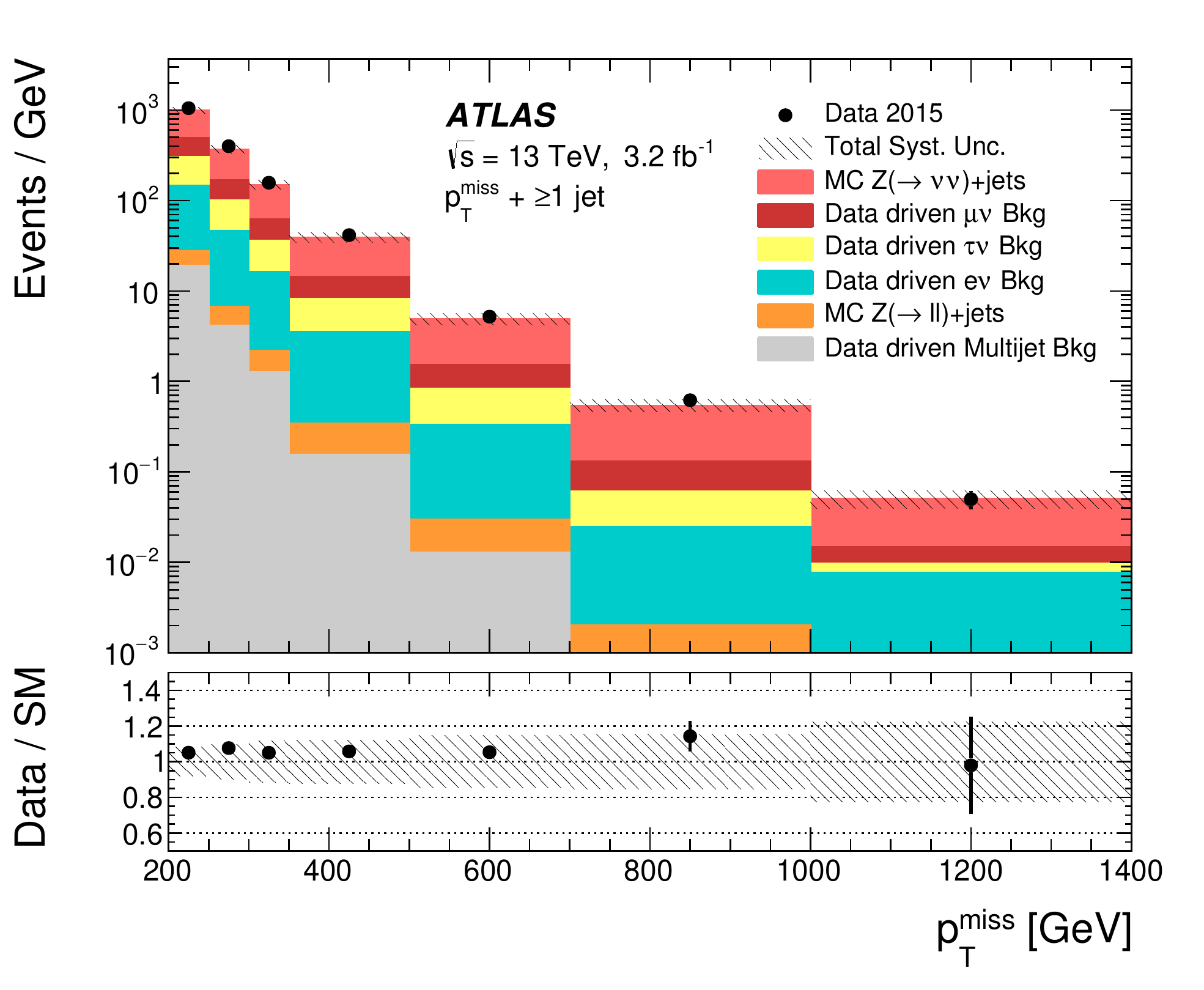}
          \label{fig:znunu-met-mono-datamc}
	}
        \subfigure[]{
          \includegraphics[width=0.47\textwidth]{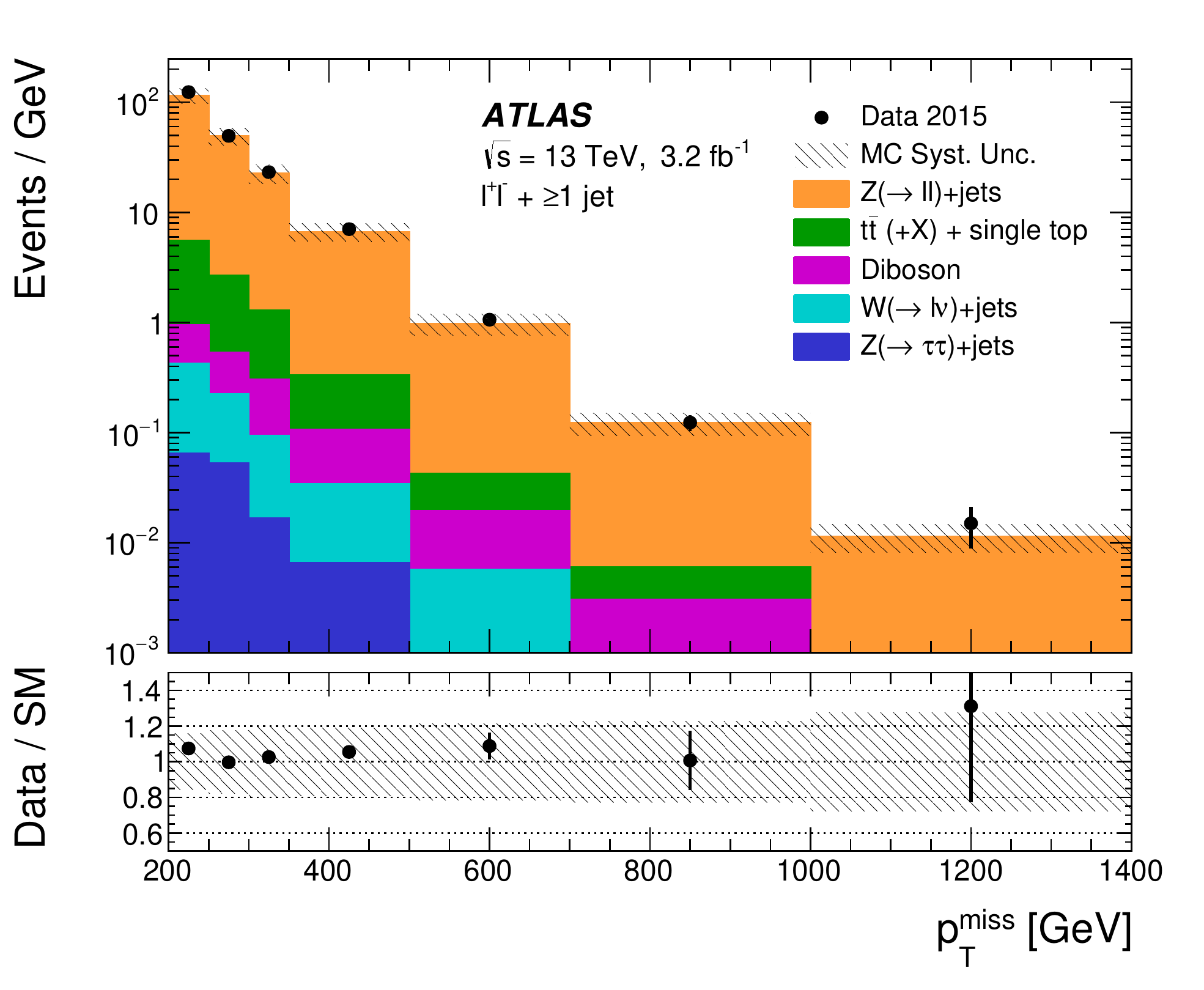}
          \label{fig:zll-met-mono-datamc}
      }
        \subfigure[]{
          \includegraphics[width=0.47\textwidth]{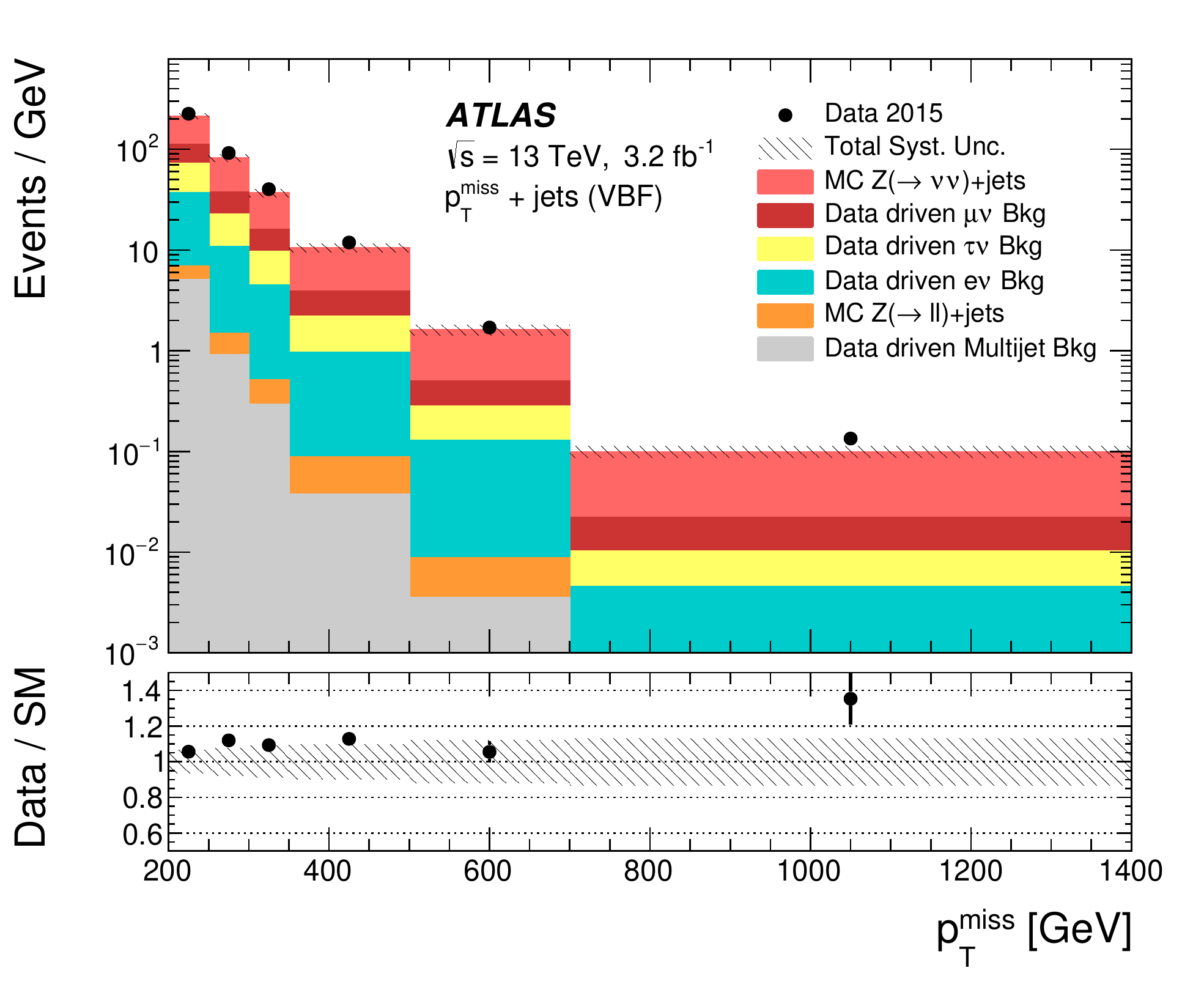}
          \label{fig:znunu-met-vbf-datamc}
	}
\subfigure[]{
          \includegraphics[width=0.47\textwidth]{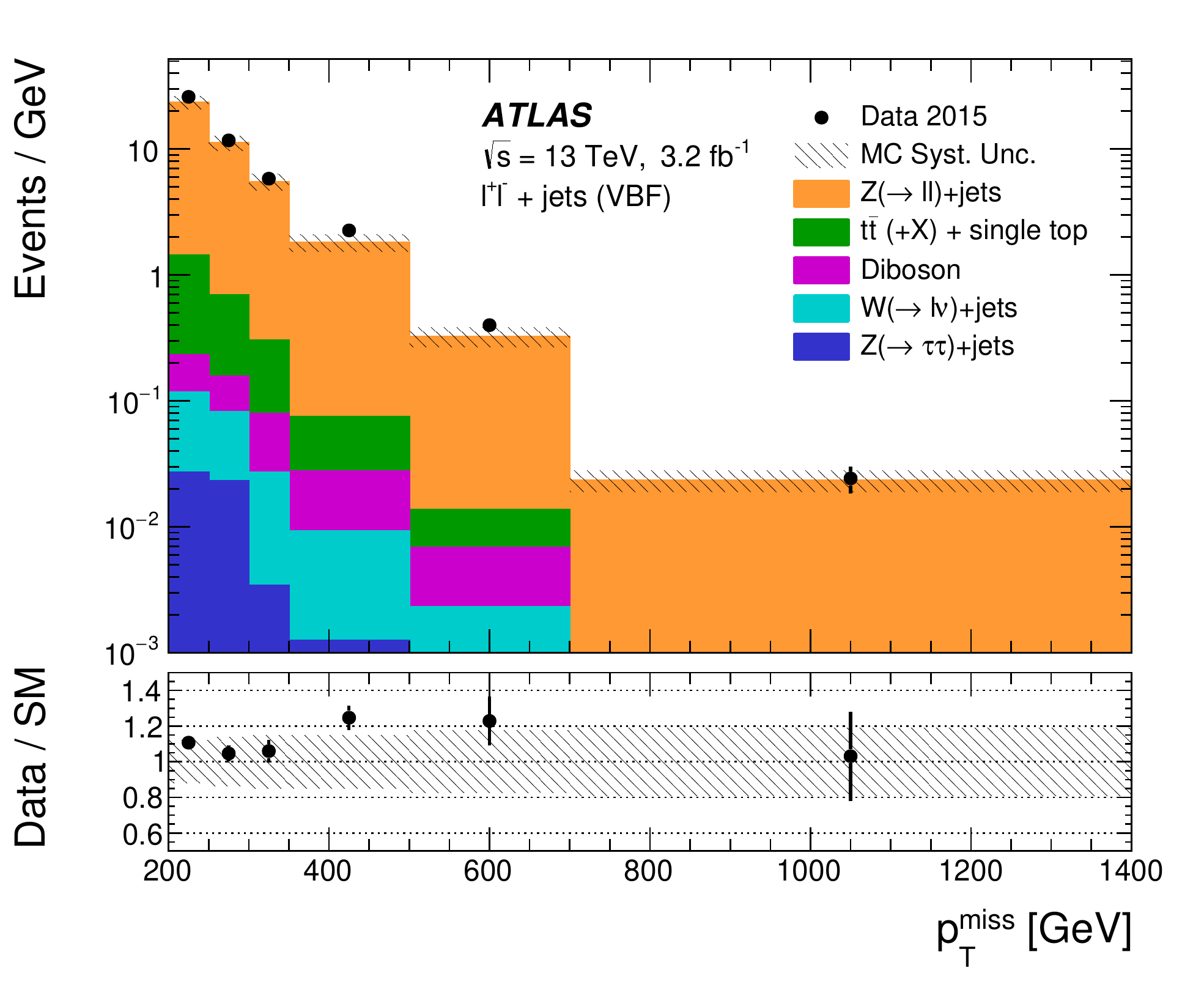}
          \label{fig:zll-met-search-datamc}
      }

	\end{center}
	\caption {Comparisons between detector-level distributions for data and MC simulation of \Znunu{} and \Zll{} events plus
          predicted backgrounds in selected (a,c) \ptmissjet{} events and (b,d) \lljet{} events as a function of the \ptmiss{}
          variable in the (a,b) \onejet{} phase space and (c,d) \vbf{} phase space. 
         The lower panel shows the ratio of data to the Standard Model prediction.
         The error bars show the statistical uncertainty of the data.
          Uncertainties in the predictions are shown as hatched bands and include
          the statistical component as well as systematic contributions from theoretical predictions, 
          lepton efficiencies and jet energy scales and resolutions to the MC predictions and uncertainties in
          the data-driven background estimates, explained in Section~\ref{sec:syst}.} 
  \label{fig:met-datamc}
\end{figure*}
\begin{figure*}
	\begin{center}
	\subfigure[]{
          \includegraphics[width=0.47\textwidth]{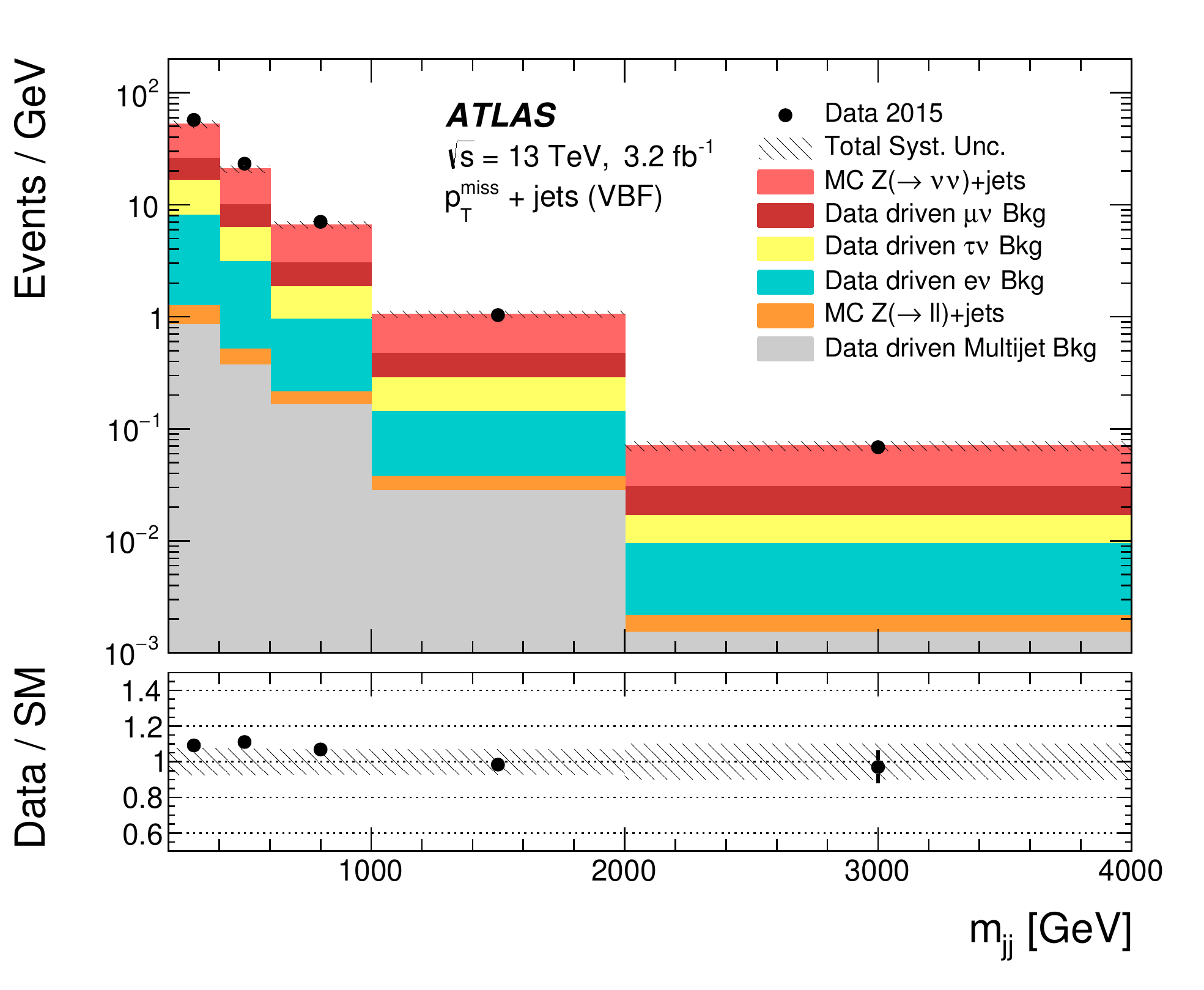}
          \label{fig:znunu-mjj-vbf-datamc}
	}
 \subfigure[]{
          \includegraphics[width=0.47\textwidth]{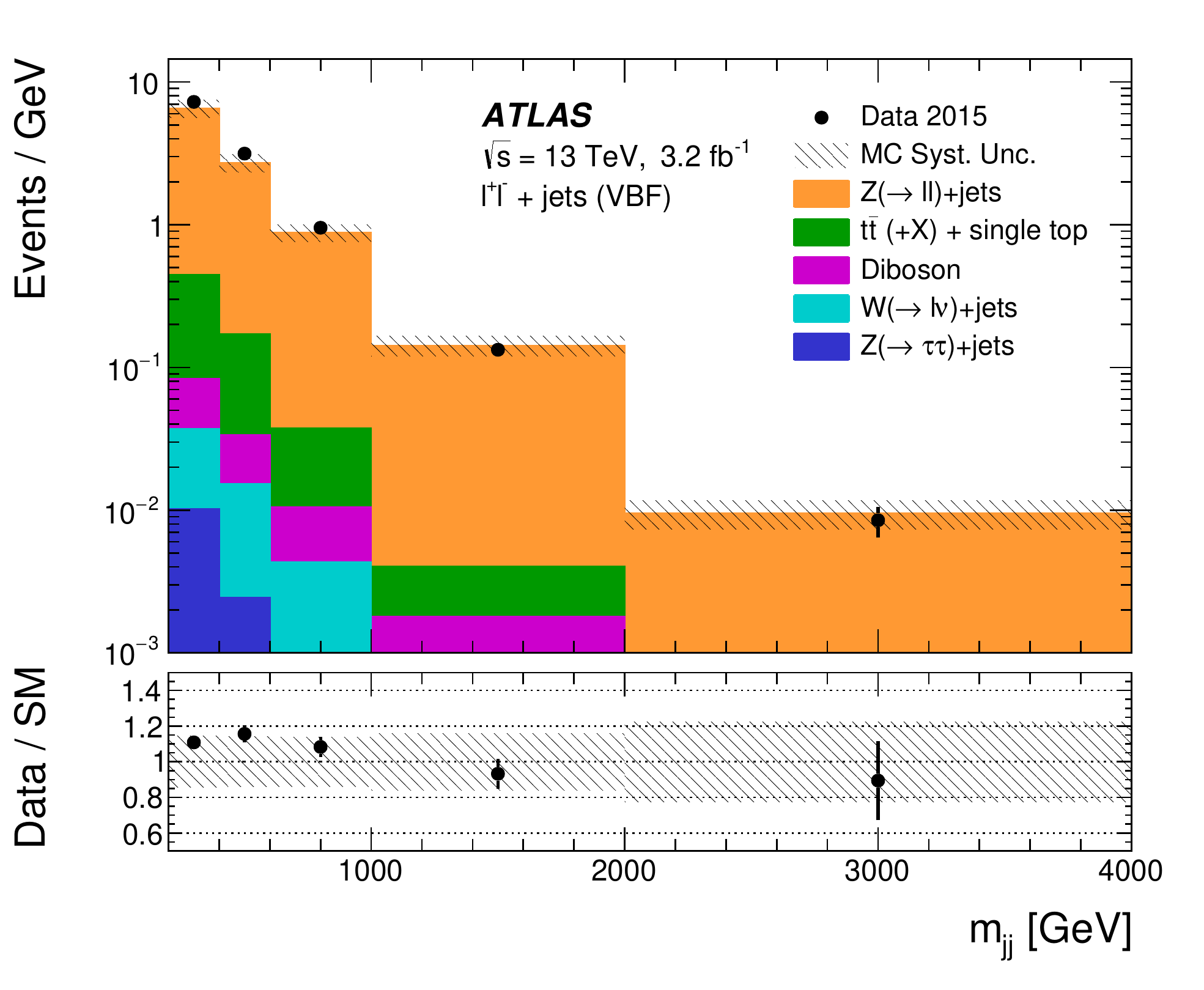}
          \label{fig:zll-mjj-vbf-datamc}
      }

        \subfigure[]{
          \includegraphics[width=0.47\textwidth]{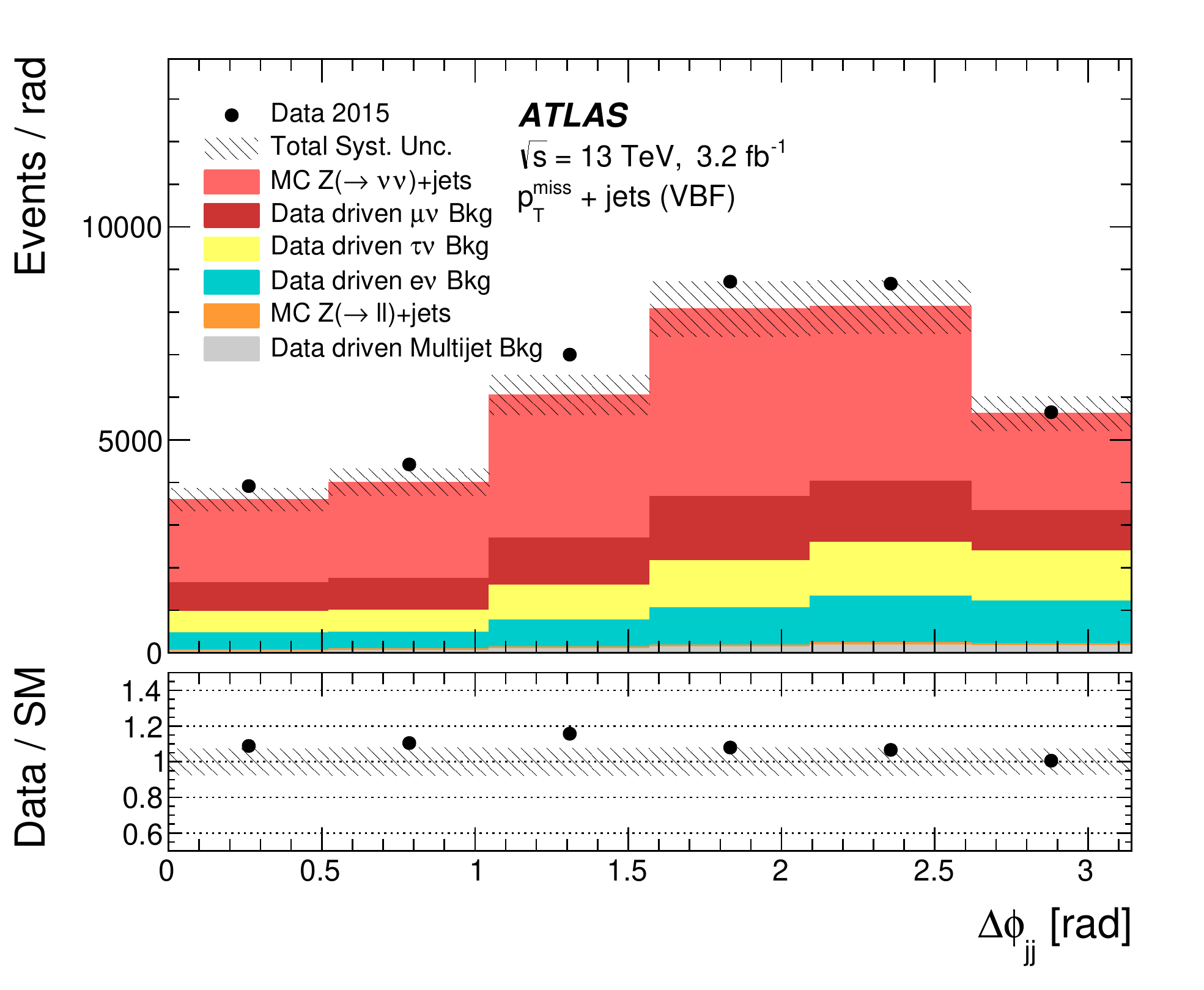}
          \label{fig:znunu-dphi-vbf-datamc}
	}
 \subfigure[]{
          \includegraphics[width=0.47\textwidth]{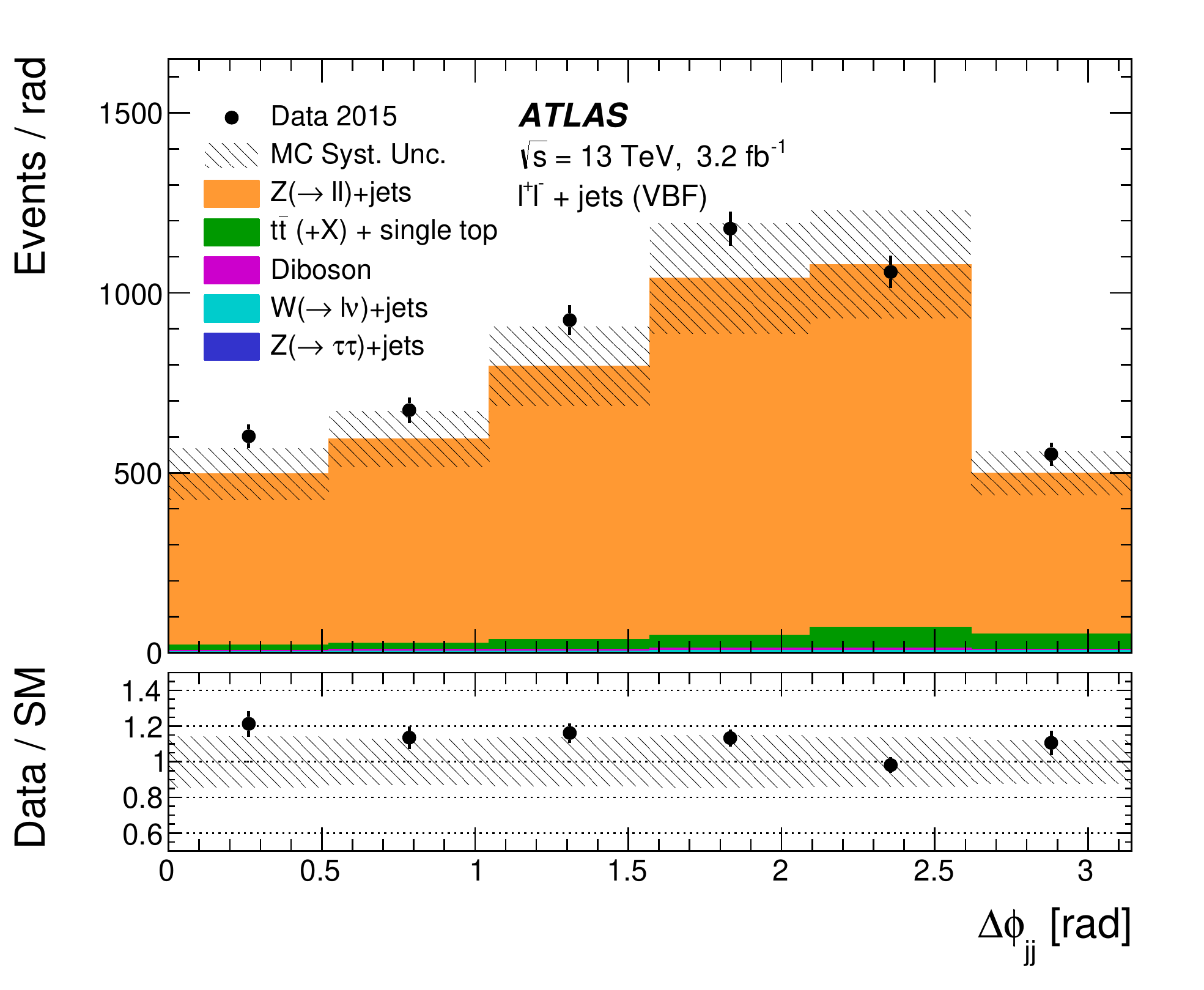}
          \label{fig:zll-dphi-vbf-datamc}
}
	\caption {Comparisons between detector-level distributions for data and MC simulation of \Znunu{} and \Zll{} events plus 
          predicted backgrounds in selected (a,c)
          \ptmissjet{} events and (b,d) \lljet{} events as a function of (a,b) \mjj{}
          and (c,d) \dphijj{} in the \vbf{} phase space. 
         The lower panel shows the ratio of data to the Standard
         Model prediction.
         The error bars show the statistical uncertainty of the data.
          Uncertainties in the predictions are shown as hatched bands and include
           the statistical component as well as systematic contributions 
          from theoretical predictions, lepton efficiencies and jet energy scales
          and resolutions to the MC predictions and uncertainties in
          the data-driven background estimates, explained in Section~\ref{sec:syst}.   }
  \label{fig:mjj-dphi-datamc}
	\end{center}
\end{figure*}
\section{Detector corrections}
\label{sec:corrections}
The data are corrected for the inefficiencies and resolutions of the
detector and trigger and are presented in terms of particle-level variables as defined in Section~\ref{sec:particle}.
Due to the similarity in the \ptmiss{} and jet selections between
numerator and denominator, corrections for the \ptmiss{} and jet-based variables arising from the jet
energy resolutions and scales
almost completely cancel in the ratio. 
Similarly, the correction factors related to the lepton veto
efficiencies cancel in the ratio.
The dominant remaining correction
factor arises from the inefficiency of reconstructing the charged leptons in
the denominator of the ratio.
The correction factor is defined as the ratio of \Rmiss{} at particle level 
to \Rmiss{} at detector level using \Znunu{} and $\Z/\gamma^{*} \to \ell^{+}\ell^{-}$ MC simulation,
in bins of the measured variables. 
The correction factor decreases with \ptmiss{} from 0.9 to 0.85 in
the muon channel and increases with \ptmiss{} from 0.7 to
0.8 in the electron channel. The number is larger for muons than
for electrons because the reconstruction efficiency for muons is higher 
for the selection criteria used in this analysis.

Event migration between bins in the distributions, due to differences in the particle-level and
detector-level variables, is small due to the relatively
wide bins and therefore ignored. In the absence of a BSM signal, dependencies of the migrations on the
underlying distributions are very similar for the numerator and
denominator and therefore systematic uncertainties arising from this
source cancel in the ratio. 
The possible impact of signals on the correction factors has been studied and found to be small.
The presence of a large BSM component in the numerator due to WIMP production with an axial-vector mediator mass 
of 1\,\TeV\ and a WIMP mass of 150\,\GeV\ (which has very different event kinematics to the SM processes) 
changes the correction factor by less than 0.5\,\%.
The injected BSM model events have a
\ptmiss{} distribution that is much harder than the \Znunu{}
contribution to the numerator, leading to changes in \Rmiss{} of 
4\,\% at low \ptmiss{} and 50\,\% at high \ptmiss{}. Such a variation
is much larger than the differences seen between data and SM
simulation.
Furthermore, injecting a Gaussian BSM contribution 
that adds events to a single bin (but remains consistent with the
data) is also found to have a very small impact; the largest change in
the correction factor is 2\,\%, in the second bin of the
\ptmiss{} distribution, which is small compared to the systematic
uncertainties. 
This test is an extreme example, where it is assumed that the full
difference between the SM prediction and data in the \Rmiss{} ratio is 
due to BSM physics in the numerator. 
It is therefore concluded that the presence of any BSM model consistent
with the data would lead to only small changes in the correction
factors and that these models can be constrained by the detector-corrected 
results. Larger BSM contributions that could cause more significant 
changes in the correction factors have already been excluded with the detector-level data.

\section{Systematic and statistical uncertainties}
\label{sec:syst}
Uncertainties in the measured detector-corrected ratios are discussed in this section and
summarised in Table~\ref{tab:systematics}.
The dominant experimental systematic uncertainties come from the
reconstruction and isolation efficiency of muons and the  reconstruction,
isolation and trigger efficiency of electrons. These uncertainties affect the detector corrections, the
\W{} background predictions from leptonic control
regions and the backgrounds to \lljet{} events.
A smaller uncertainty in the $\tau$ reconstruction efficiency, affecting the 
$\tau$ veto, is also included.
These are collectively labelled ``Lepton efficiency'' in the table.
Uncertainties in the jet energy scale and resolution, labelled ``Jets''
in the table, affect the
background predictions as well as the detector
corrections. The latter arises due to small differences between the selected
events for the numerator and denominator, such as the removal of jets close to leptons.
The uncertainty from the difference in the choice of control region for the \Wtaunu{}
background prediction, described in Section~\ref{sec:backgrounds}, is also included.
For the multijet background
estimation a 50\,\% uncertainty in the number of predicted 
events, together with a smaller uncertainty found by varying the
selection criteria for events used as input for the smearing
method, is assumed.
The difference between the reweighted and nominal MC simulation background
prediction of \lljet{} events is taken as an uncertainty. The
reweighting factor is obtained from an $e^{\pm}\mu{\mp}$ control
region, described in Section~\ref{sec:backgrounds}.
Statistical uncertainties from the finite size of the MC simulation samples used to determine
the detector corrections, in the \W{} control region data, and MC 
simulation samples used for extrapolations are also included.

Three categories of theoretical uncertainties are considered.
Firstly, an uncertainty of 30\,\% in the cross-section of processes
involving top quarks in the numerator is assigned.
This indirectly affects the extrapolation of \W{} events to the signal
region by altering the number of top quark events in the control
regions. 
The uncertainty value is motivated by top-quark-enhanced control
regions constructed using the same criteria as the \W{} control regions but in addition requiring
either one or two jets consistent with containing a $b$-hadron.
Discrepancies between MC simulation and data of up to 30\,\% are seen in these control regions,
which justifies the large uncertainty.
Secondly, theoretical uncertainties that affect the
extrapolations between the control and signal regions for \W{}
backgrounds are included. These are estimated by varying the factorisation,
renormalisation, resummation scales (each scale varied by factors of 0.5 and 2) and
the CKKW matching~\cite{Catani:2001cc,Hoeche:2009rj}  scale between 30\,\GeV\ and 15\,\GeV\ (the nominal being 20\,\GeV).
These variations were found to affect the control and signal regions in the same way and the resulting uncertainties 
are therefore treated as fully correlated between the two.
PDF uncertainties are derived for the nominal NNPDF3.0nnlo PDF set~\cite{Ball:2014uwa} as well as
the MMHT2014~\cite{Harland-Lang:2014zoa} and CT14~\cite{Dulat:2015mca} PDF sets using their recommended PDF uncertainty prescription.
A combined PDF uncertainty is then obtained from the envelope of the three PDF families and their respective uncertainties.
An uncertainty from the strong coupling constant
$\alpha_{\mathrm{S}} \left(m_{\mathrm{Z}}\right)$ is
derived using up and down variations to 0.117 and 0.119, respectively
(the nominal value being 0.118). 
Thirdly, the change in the \W{} background predictions when using \sherpa{}~\cite{Gleisberg:2008ta}
v2.1.1 (which uses the CT10nlo~\cite{Lai:2010vv}  PDF set and has some technical
differences in the parton shower compared to v2.2.0) or \mgfive{} v2.2.2~\cite{Alwall:2011uj} instead of \sherpa{} v2.2.0 is considered.
The second and third theoretical sources are included as
``$W$ theory'' in Table~\ref{tab:systematics}.
The correction factors do not change significantly when varying the SM MC event generator.

For each of the three data samples (\ptmissjet{}, \eejet{} and \mmjet{}), the statistical uncertainty is taken as the 
Poisson error. 
For bins containing  a small number of events,
this uncertainty in the denominator leads to an asymmetric uncertainty
in the ratio. 
Table~\ref{tab:systematics} summarises the size of each  systematic
uncertainty and the statistical uncertainty from the data for the lowest and highest \ptmiss{} bins in the
\onejet{} phase space and the lowest and highest \mjj{} bins in the
\vbf{} phase space of the combined ratio.
The uncertainties vary monotonically as a function of the respective observable.

\begin{table}
\centering
\caption{\label{tab:systematics} Summary of the uncertainties in the
  measured ratio \Rmiss{} for the lowest and highest \ptmiss{} bins in the
  \onejet{} phase space and the lowest and highest \mjj{} bins in the
  \vbf{} phase space. The statistical uncertainty is from
  the data. Statistical uncertainties in the MC simulation
  are included as systematic uncertainties.
  The uncertainties vary monotonically as a function of the respective observable.}
\begin{tabular}{ l|rrrr} 
 \hline
 \textbf{Systematic uncertainty source}     & Low \ptmiss{} [\%] & High \ptmiss{} [\%]  &
 Low \mjj{} [\%] & High \mjj{} [\%] \\ 
\hline
Lepton efficiency   & $+3.5$,  $-3.5$  & $+7.6$, $-7.1$ & $+3.7$,  $-3.6$  & $+4.6$, $-4.4$\\
Jets                        & $+0.8$,  $-0.7$  & $+2.2$, $-2.8$  &  $+1.1$,  $-1.0$  & $+9.0$, $-0.5$ \\
\Wtaunu{} from control region         & $+1.2$,  $-1.2$  & $+4.6$, $-4.6$  & $+1.3$,  $-1.3$  & $+3.9$, $-3.9$  \\
Multijet                  &  $+1.8$,  $-1.8$  & $+0.9$, $-0.9$  &   $+1.4$,  $-1.4$  & $+2.5$, $-2.5$\\
Correction factor statistical    & $+0.2$,  $-0.2$  & $+2.0$, $-1.9$ & $+0.4$,  $-0.4$  & $+3.8$, $-3.6$  \\
\W{}  statistical            & $+0.5$,  $-0.5$  & $+24$, $-24$   &  $+1.1$,  $-1.1$  & $+6.8$, $-6.8$ \\
\W{} theory         & $+2.4$,  $-2.3$  & $+6.0$, $-2.3$  & $+3.1$,  $-3.0$  & $+4.9$, $-5.1$ \\
Top cross-section  & $+1.5$,  $-1.8$  & $+1.3$, $-0.1$  &  $+1.1$,  $-1.2$  & $+0.5$, $-0.4$   \\
$Z\rightarrow \ell\ell$ backgrounds & $+0.9$,  $-0.8$  & $+1.1$, $-1.1$ & $+1.0$,  $-1.0$  & $+0.1$, $-0.1$ \\ 
\hline
\textbf{Total systematic uncertainty}               & $+5.2$, $-5.2$ & $+27$, $-26$   & $+5.6$, $-5.5$ & $+14$, $-11$   \\
\hline
\textbf{Statistical uncertainty}                   & $+1.7$, $-1.7$ & $+83$, $-44$   &  $+3.5$, $-3.4$ & $+35$, $-25$  \\    
\hline
 \textbf{Total uncertainty}   & $+5.5$, $-5.4$ & $+87$, $-51$  & $+6.6$, $-6.5$ & $+38$, $-27$  \\    
\hline
\end{tabular}
\end{table}

\section{Combination}
\label{sec:combo}
After subtracting the estimated backgrounds from the selected \lljet{}
event sample in the data, and applying the bin-by-bin detector correction factor, the electron
and muon denominators are combined
using the best linear unbiased estimate (BLUE) combination method~\cite{Nisius:2014wua}, which takes
into account the relative precision of the two measurements.
The technique correlates 
the statistical and systematic uncertainties between the two measurements 
and between all bins in all distributions. 
The combined result produces an average for  \lljet{} of one flavour
in the denominator. 
The combination is iterated once,
replacing the statistical uncertainty in the observed number of \Zmumu{} and
\Zee{} events with that obtained from the expected number of  events after the first combination. 
This removes the effect of undue weight being given to the channel in which the number
of events has fluctuated down.
In the combination, statistical correlations
between bins are accounted for using a bootstrap
method~\cite{PhysRevD.39.274}.
The  \Zll{} background uncertainty is assumed to be fully correlated
or anti-correlated between bins, depending on whether the fit to
estimate \Zll{} background events increases or decreases the result
from MC simulation in a given bin. The correlation between bins for the electron
and muon efficiency uncertainties is found by considering the separate
sources that contribute to the total uncertainties.
All other sources of systematic uncertainty are assumed to be fully correlated across bins in the combination. 
The $p$-value for the compatibility of the two channels for all four
distributions is 74\,\%.
The ratio is then formed by subtracting the estimated backgrounds from the selected
\ptmissjet{} event sample in the data and dividing by the combined denominator.
Again, each source of systematic uncertainty is assumed
to be fully correlated between numerator and denominator.
A cross-check using a maximum-likelihood fitting method gives
consistent results.

\section{Results}
\label{sec:results}
Figure~\ref{fig:final-combined-ratios} shows the four combined differential
measurements of \Rmiss{} compared to the average of the \sherpa{} v2.2.0 SM particle-level predictions 
for the muon and electron channels. The measurement is consistent with the SM prediction within
statistical uncertainties. 
The uncertainty in the SM prediction,
found from the factorisation and renormalisation scale variations as
well as the NNPDF3.0nnlo PDF uncertainty, explained in Section~\ref{sec:syst}, is
shown as a red hatched band in the figure.
The SM predictions do not include NLO electroweak corrections beyond
final-state photon radiation. These corrections were studied in
Ref.~\cite{Denner:2012ts} for the \Z{} boson production at a 
centre-of-mass energy of 8\,\TeV\ and are very similar for the numerator
and denominator with a residual effect of up to 1\,\% on the ratio. 

\begin{figure}[htbp]
        \begin{center}
          \subfigure[]{\includegraphics[width=0.48\textwidth]{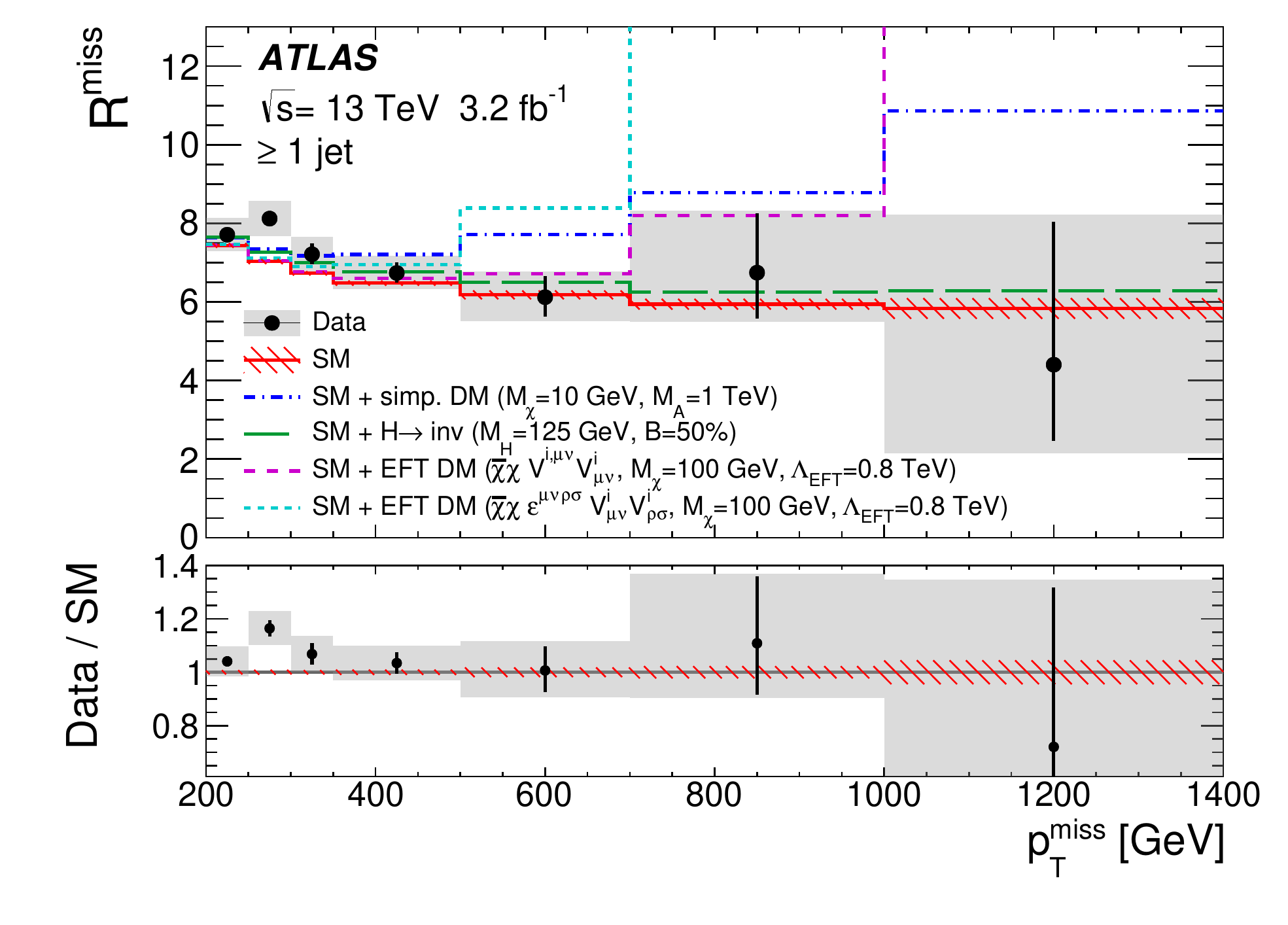} \label{fig:ratio-met-mono}}
          \subfigure[]{\includegraphics[width=0.48\textwidth]{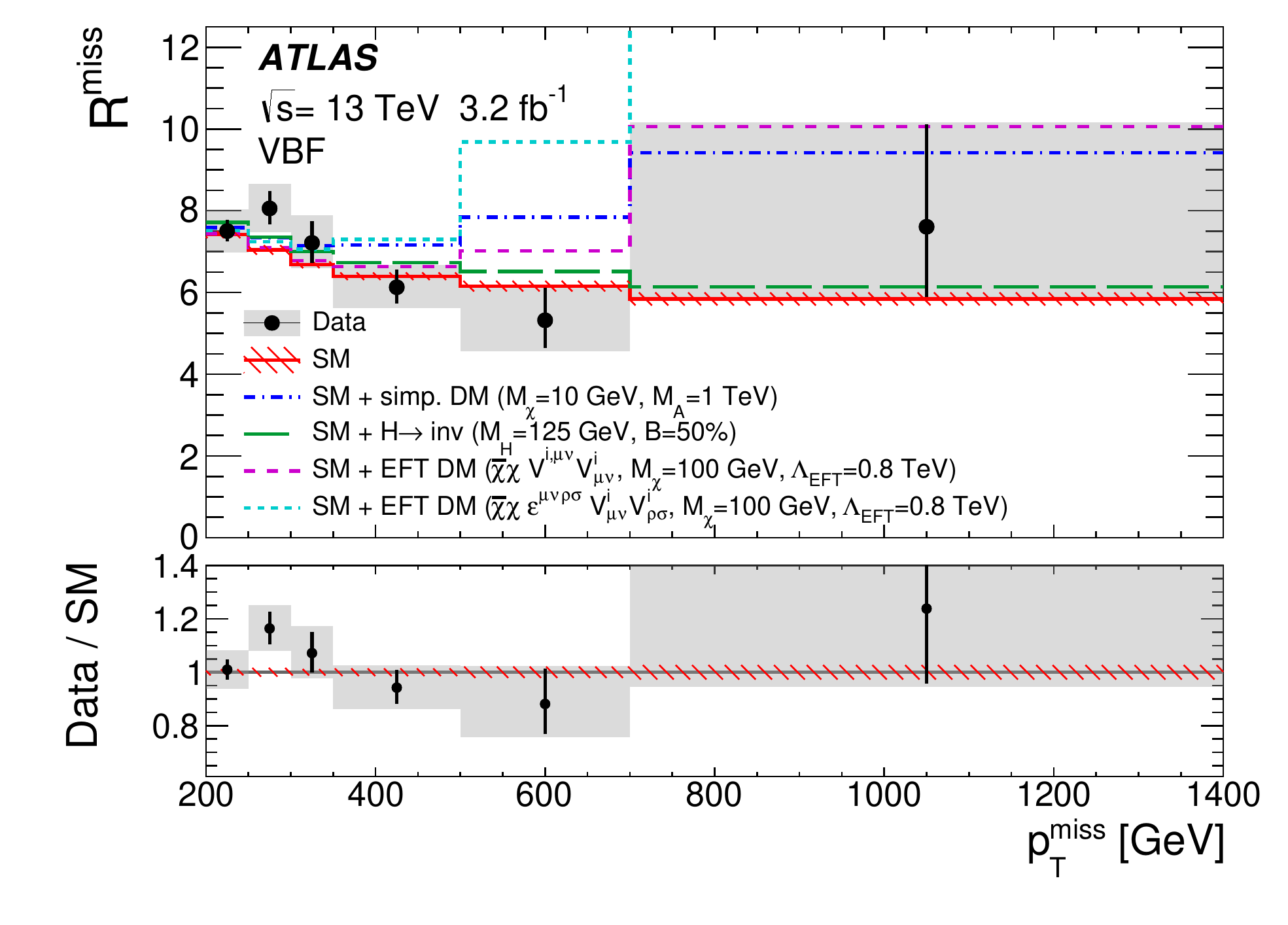}           \label{fig:ratio-met-vbf}}

          \subfigure[]{\includegraphics[width=0.48\textwidth]{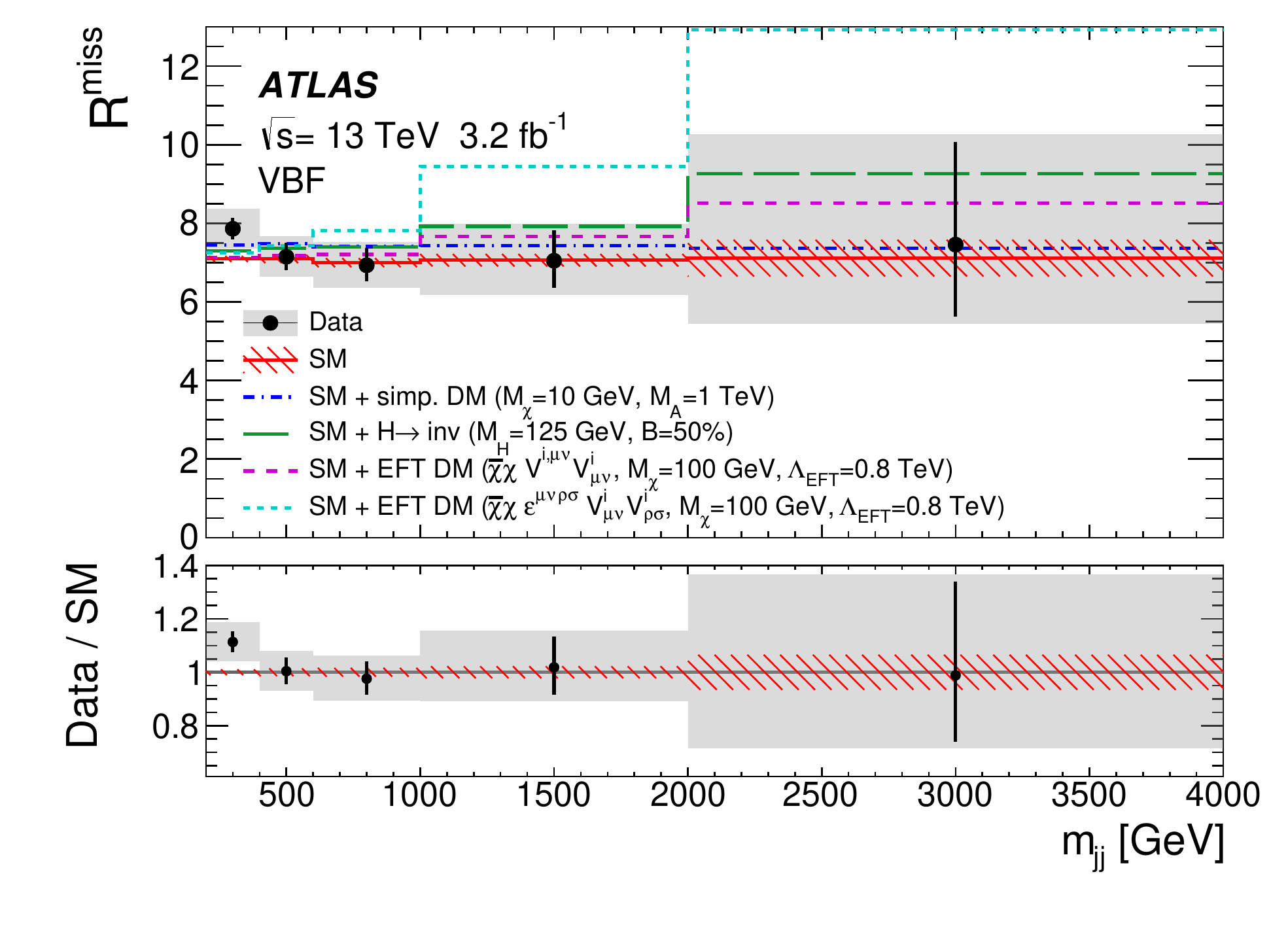}}
          \subfigure[]{\includegraphics[width=0.48\textwidth]{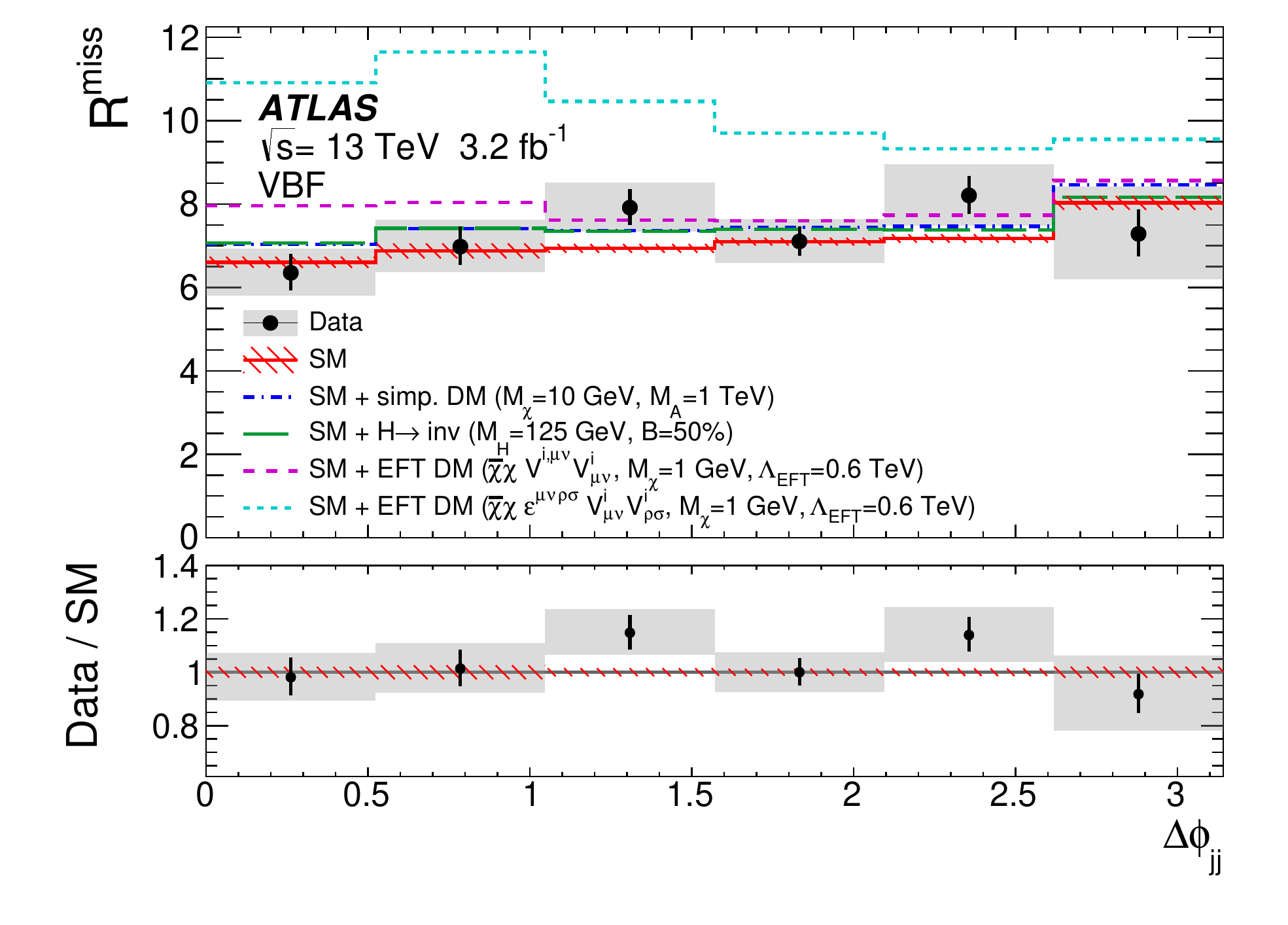}}
        \end{center}
        \caption {Measured \Rmiss{} as a function of (a)
          \ptmiss{} in the \onejet{} region, (b) \ptmiss{} in the \vbf{}
          region, (c) \mjj{} in the \vbf{} region and (d)  \dphijj{} in
          the \vbf{} region. Statistical uncertainties are shown as
          error bars and the total statistical and systematic uncertainties
          are shown as solid grey bands. The results are compared to
          the SM prediction and to SM+BSM for
          four BSM models. One is a simplified model of WIMP production with 
          an $s$-channel exchange of an axial-vector mediator with mass of 1\,\TeV\ coupling to quarks and a
         WIMPs with a mass of 10\,\GeV, another represents the Higgs boson decaying to invisible
         particles with 50\,\% branching fraction, and another two represent the predictions of two EFT operators 
         allowing the production of WIMP dark matter through
         interactions with vector bosons (with differing charge-parity properties in the interaction). The \Rmiss{} values 
         of the third and fourth models in the highest \ptmiss{} bin in the \onejet{} region are 18.8 and 38.3, respectively,
         and in the highest \ptmiss{} bin in the \vbf{} region the fourth model has an \Rmiss{} value of 19.4. 
         The red hatched error bars correspond to the uncertainty in the SM prediction.
The bottom panel shows the ratio of data to the SM prediction.}
  \label{fig:final-combined-ratios}
\end{figure}

Also shown in the Figure~\ref{fig:final-combined-ratios} is a comparison with SM+BSM for
four BSM models. These four models comprise a simplified model for WIMP production with 
an $s$-channel exchange of an axial-vector mediator with a mass of 1\,\TeV\ and a
WIMP mass of 10\,\GeV, a Higgs boson decaying to invisible particles with 50\,\% branching fraction, 
and two examples of effective field theory operators 
(each with different charge-parity properties) involving couplings of WIMP dark-matter candidates
with vector bosons. These models are described in Section~\ref{sec:MC}.

\section{Discussion}
\label{sec:discuss}

In Figures~\ref{fig:ratio-met-mono} and~\ref{fig:ratio-met-vbf}, both the measurements and the SM predictions 
show a ratio \Rmiss{} of approximately 7.5 at $\ptmiss{} = 200$\,\GeV, decreasing with \ptmiss{} to approximately 6,
which is very close to the SM ratio of branching fractions in the numerator and denominator 
of 5.9~\cite{Olive:2016xmw}.\footnote{The denominator also includes the presence of 
the $\gamma^*$ mediator, which is not present in the numerator and would influence 
the $\Rmiss$ ratio in the SM; however, this contribution is small as the dilepton invariant 
mass is required to be close to the \Z{} mass.} 
The ratio is larger at lower \ptmiss{} values
due to the fiducial requirements on the charged leptons in the
denominator. 
At higher \ptmiss{} values the leptons are more 
central and have larger \pt{}, and are therefore more likely to pass
the fiducial requirements. The removal of jets overlapping with
charged leptons, described in Section~\ref{sec:particle}, is only
relevant to the denominator. In particular, a slight increase in the ratio towards large
\dphijj{} values is seen, indicating that jets with this topology are more likely to be
removed in the denominator.
The data and SM predictions are in agreement with an overall $p$-value including all distributions of 22\,\%
taking into account statistical and systematic correlations. 
In addition to
the measured ratios, a covariance matrix for all four
distributions, taking into account the statistical and systematic correlations
between all bins in the data, is produced using a bootstrap procedure. When forming the covariance
matrix the uncertainties are symmetrised by taking the maximum of the
upward and downward uncertainties.

The detector-corrected ratio for all four distributions, together with the covariance matrix for the statistical and systematic uncertainties, 
as well as model uncertainties in the SM prediction for the numerator
and denominator, and acceptance uncertainties in the WIMP model,
are used to set limits on the mass of the axial-vector mediator ($m_A$) and WIMP candidate ($m_\chi$).
Factors affecting the WIMP model signal acceptance include uncertainties in the modelling of initial- and
final-state radiation in simulated samples, uncertainties in 
PDFs and the choice of $\alpha_{\mathrm{S}}\left(m_{\mathrm{Z}}\right)$, and the choice of renormalisation and factorisation scales.

Limits on dark-matter production models are set by first constructing the $\chi^2$ function
\begin{equation*}
\chi^2 = {(\mathbf{y}_\mathrm{data} - \mathbf{y}_\mathrm{pred})}^{T} C^{-1} (\mathbf{y}_\mathrm{data} - \mathbf{y}_\mathrm{pred}),
\end{equation*}
where $\mathbf{y}_\mathrm{data}$ and $\mathbf{y}_\mathrm{pred}$ are the vectors of the measured \Rmiss{} values 
and the predicted \Rmiss{} values for the hypothesis under test across the four distributions under study, $C$
is the total covariance matrix defined as the sum of the statistical, experimental systematic and theoretical systematic covariances. 
The \cls technique\,\cite{Read:2002hq,Junk:1999kv} evaluated using the asymptotic 
approximation\,\cite{Cowan:2010js} is used to derive upper limits.

The overall rate and kinematic properties of events in the axial-vector mediator WIMP model 
under study are defined by four parameters: the WIMP candidate mass, the mediator mass and 
the strengths of the mediator interaction with quarks and WIMPs.
The expected and observed 95\,\% confidence level (CL) exclusion limits as a function of mediator 
and WIMP mass are shown in Figure~\ref{fig:exclusions-dma}, for fixed mediator couplings of $g_q = 0.25$ and $g_\chi = 1$.
Expected limits are shown with $\pm 1\sigma$ bands indicating the range of the expected limit 
in the absence of a signal. Observed limits are shown with a band including the effect of 
$\pm 1\sigma$ theoretical uncertainties in the WIMP model cross-section.
Also highlighted is the region where perturbative unitarity is violated (where 
$m_\chi>\sqrt{\pi/2}\,m_A$)\,\cite{Kahlhoefer:2015bea}. The points in the mass plane compatible with the
relic density measured by Planck\,\cite{Adam:2015rua} and
WMAP\,\cite{WMAP} are represented by a red continuous line, with WIMP
masses below this line or mediator masses to the right of this
line corresponding to dark-matter overproduction.  
The highest mediator mass observed (expected) to be excluded at 95\,\% CL is 1.24\,\TeV\ (1.09\,\TeV).
For comparison, limits set using detector-level observables~\cite{EXOT-2015-03} are also shown. 
For high mediator masses, the expected limits in the present analysis are slightly weaker, 
due to the limited number of events in the denominator, whereas the observed limits are slightly stronger
compared to the detector-level analysis.
This difference between expected and observed limits is driven entirely by systematic uncertainty correlations 
between bins of the corrected distributions.
Switching between using the default correlation model and a simple correlation model
assuming 100\,\% correlation between bins for each source of experimental systematic uncertainty changes the observed limit in
mediator mass by approximately 10\,\GeV.
The measurements presented in this paper 
have enhanced sensitivity to models with large WIMP masses and low mediator masses, with respect to the detector-level analysis
presented in Ref.~\cite{EXOT-2015-03}, 
due to the use of a larger fiducial volume and the use of differential information with associated correlations.

\begin{figure}
          \includegraphics[width=0.95\textwidth]{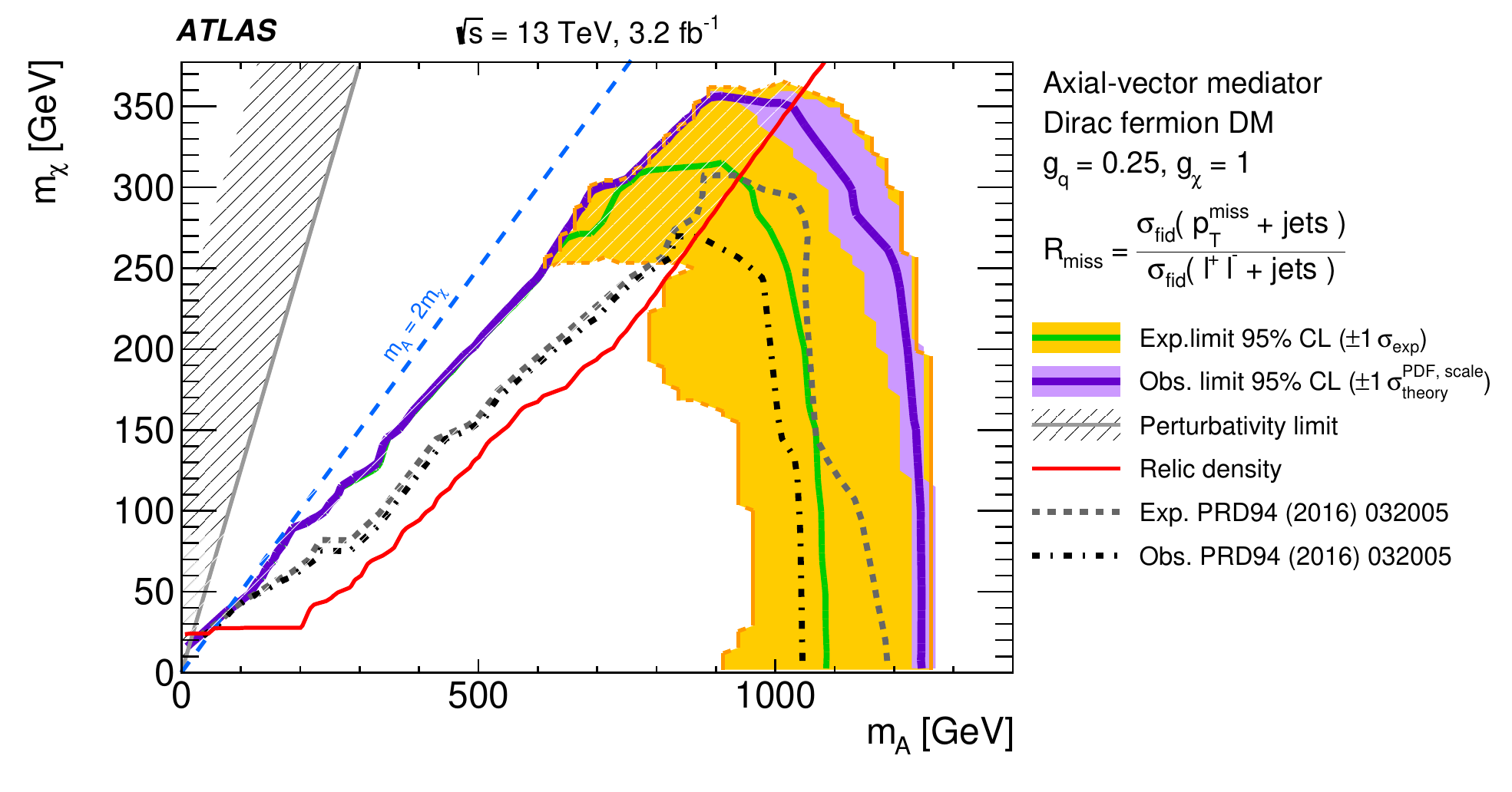} 
        \caption {Exclusion contours (at 95\,\% CL) in the WIMP--mediator
          mass plane for a simplified model with an axial-vector
          mediator and couplings $g_q = 0.25$ and $g_{\chi}=1$.
The solid purple (green) curve shows the the observed
(expected) limit. The yellow filled region around the expected limit indicates the effect of 
$\pm 1\sigma$ experimental uncertainties in the expected limit.
The red curve corresponds to the expected relic
density. The grey hatched region shows the region of non-perturbativity
defined by WIMP mass greater than $\sqrt{\pi/2}$ times the mediator
mass.
Also shown, for comparison, are limits set using detector-level event
counts from Ref.~\cite{EXOT-2015-03}.
The exclusion is based on the global fit to the \ptmiss{}
distributions in the \onejet{} and \vbf{} phase spaces, and the \mjj{}
and \dphijj{} distributions in the \vbf{} phase space.
}
  \label{fig:exclusions-dma}
\end{figure}

The detector-corrected data are also used to search for Higgs boson decays to invisible particles in the same manner.
Limits are placed on the production rate of the Higgs boson multiplied by its branching fraction to invisible particles
relative to the total Higgs boson production rate as predicted by the SM~\cite{deFlorian:2016spz}.
The expected 95\,\% CL upper limit for a Higgs boson with a mass of 125\,\GeV\ is found to be
0.59
with a range of
[0.47, 1.13]
from $\pm 1\sigma$ experimental uncertainties.
The observed upper limit at 95\,\% CL is
0.46.
The most important distribution for setting limits in this model is
\mjj{}, although some additional expected sensitivity is achieved
from \dphijj{}. 
The observed limits are stronger than expected due to systematic uncertainty correlations between bins in the corrected ratios. 
This is to be compared with an exclusion limit of 0.28 (0.31 expected) at 95\,\% CL using a 20~\ifb{}
8\,\TeV\ data set~\cite{HIGG-2013-16}, with an event selection optimised for this particular process.

The detector-corrected data are further used to set limits on the production of Dirac-fermion dark matter 
in a generalised effective field theory (EFT) where dark matter interacts only with electroweak bosons.
Limits are set as a function of the invariant mass of the dark-matter candidate and the EFT scale, $\Lambda$, 
which can be related to a UV-complete model by the relationship
$1/\Lambda^2 \sim g_\mathrm{SM}\, g_\chi /M^2$ where $g_\mathrm{SM}$ and $g_\chi$ would be couplings 
of the SM and dark-matter particles to some hypothetical heavy mediating particle with mass $M$.
The scenario where production is dominated by two specific dimension-seven effective operators, 
$\bar{\chi}\chi V^{\mu\nu}V_{\mu\nu}$ and $\bar{\chi}\chi \varepsilon^{\mu\nu\rho\sigma} V_{\mu\nu}V_{\rho\sigma}$,
with differing $CP$ properties in the interaction between two
electroweak bosons ($V=W/Z$) and two dark-matter particles is considered. 
This EFT is described in Ref.~\cite{Cotta:2012nj} where an assessment of the EFT validity 
for these operators is also conducted. These operators are particularly interesting as sensitivity 
benchmarks since they are insensitive to constraints from $Z$-boson invisible-width measurements.

Figure~\ref{fig:exclusions-vbfeft} shows the 95\,\% CL expected and observed limits extracted from the 
fit to all four measured distributions, compared to indirect-detection limits. 
For the $CP$-conserving operator, expected (observed) limits on the EFT scale
range from 0.78 (0.89)\,\TeV\ at low ($<200$\,\GeV) dark-matter mass to 0.61 (0.71)~TeV\ at dark-matter masses of 1\,\TeV. 
Limits for the $CP$-violating operator are stronger than for the $CP$-conserving equivalent, 
ranging from 0.99 (1.14)\,\TeV\ at low dark-matter masses to 0.77 (0.89)\,\TeV\ at dark-matter masses of 1\,\TeV.
Limits from indirect dark matter detection experiment results~\cite{Ackermann:2011wa,Aliu:2012ga,Cotta:2012nj} 
interpreted in terms of these effective operators overlaid on Figure~\ref{fig:exclusions-vbfeft} are sensitive 
up to EFT scales of 100--200\,\GeV.

\begin{figure}[thbp]
\centering
 \includegraphics[width=0.65\textwidth]{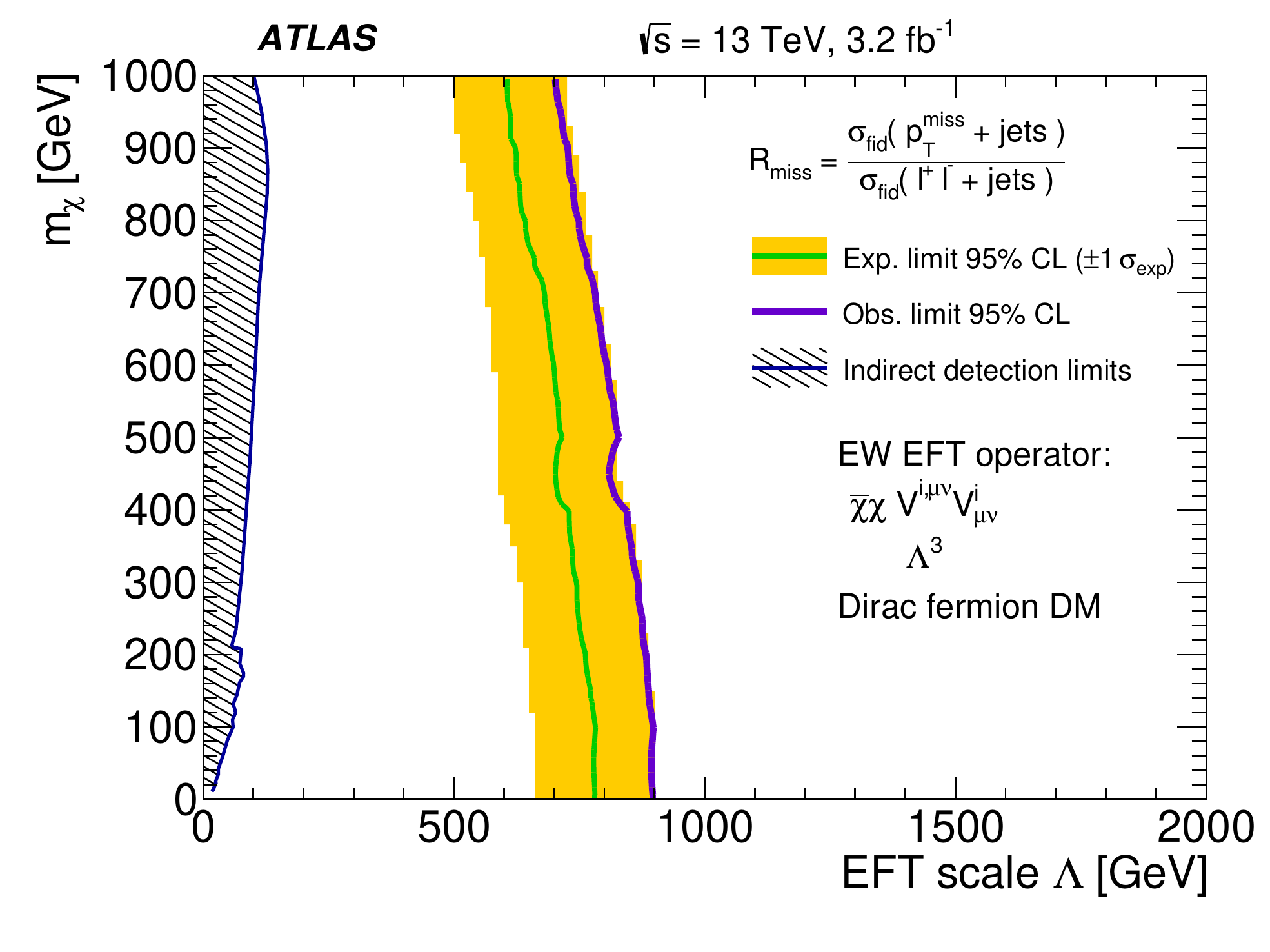}
 \includegraphics[width=0.65\textwidth]{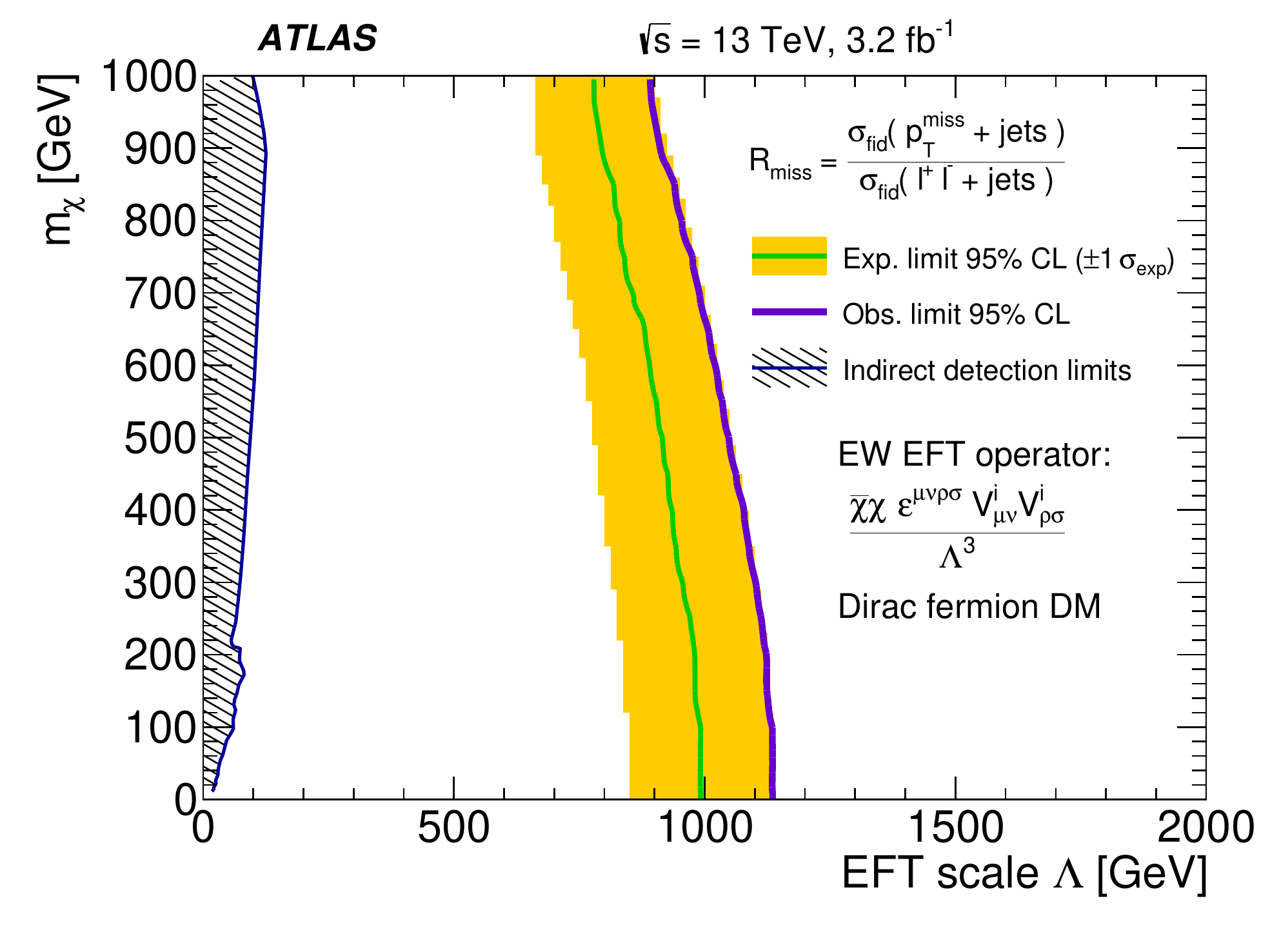}
\caption{Exclusion contours (at 95\,\% CL) for Dirac-fermion dark matter produced via a contact interaction with two electroweak bosons
as described in an effective field theory with two dimension-seven operators (described in text) with different charge-parity properties.
Limits are set as a function of dark-matter mass and the effective field theory scale, $\Lambda$.
The solid purple (green) curve shows the median of the observed (expected) limit. 
Also shown are limits on these operators from indirect-detection experiments.
The yellow filled region around the expected limit indicates the effect of 
$\pm 1\sigma$ experimental uncertainties in the expected limit.
The exclusion is based on the global fit to the \ptmiss{}
distributions in the \onejet{} and \vbf{} phase spaces, and the \mjj{}
and \dphijj{} distributions in the \vbf{} phase space.
}
\label{fig:exclusions-vbfeft}
\end{figure}

The limits presented above assume a single operator would dominate the dark-matter production rate, 
but the detector-corrected data and covariance information can be used to explore
more complex scenarios where multiple operators could contribute to the observed production rate with arbitrary 
relative rates and induce interference contributions between processes that would 
introduce non-trivial shapes and correlations between all three observables presented in this paper.
The impact on the ratios in such an EFT model is demonstrated in Figure~\ref{fig:final-combined-ratios} and is unlike the
axial-vector mediator WIMP model and Higgs model presented above which predominantly modify 
only the \ptmiss{} and \mjj{} distribution shapes, respectively.

The data have been corrected for detector effects and can be compared
to any SM prediction or a combination of SM and BSM predictions at particle level, where the BSM
model produces \ptmissjet{} final states. 
Models that also produce final states with at least one prompt lepton and
\ptmiss{} cannot be accurately compared to the data. This is because they will have
been included in the \W{} background estimation, for which the
extrapolation factors from control regions to the signal regions, determined
using SM MC simulation, would be incorrect.
Similarly, new-physics models with two leptons, entering the
denominator, can only be reliably constrained by the data if the
leptons have kinematics that are qualitatively similar to those in SM events,
otherwise differences in the lepton efficiency correction factors may
be observed.
The data, together with the full covariance matrix for the uncertainties, are
stored in \hepdata{}~\cite{Buckley:2010jn} and the analysis
is included as a routine in the \rivet~\cite{Buckley:2010ar} software
framework, in
order to ease comparisons. Also stored in \hepdata{} are the SM
numerator and denominator predicted by \sherpa{}, together with the covariance matrix for
their uncertainties, such that these can be used when comparing to BSM
models without having to simulate the SM contributions.

\section{Conclusions}
\label{sec:conclusions}
Observables sensitive to the anomalous production of events containing one or more
hadronic jets with high transverse momentum produced in
association with a large \ptmiss{} have been measured differentially
with respect to a number of properties of the hadronic system.
The results are
presented as a measurement of the ratio of \ptmissjet{} to \lljet{}
events and are fully corrected for detector effects.
This is the first detector-corrected measurement of observables 
specifically designed to be sensitive to dark-matter production.

The analysis uses 3.2\,\ifb\ of proton--proton collision data 
recorded by the ATLAS experiment at the LHC 
at a centre-of-mass energy of 13\,\TeV.
The results are presented in two phase-space regions defined by the hadronic system:
a \onejet{}  inclusive sample and a \vbf{} topology. 
The particle-level differential ratio measurements are found to be
consistent with the SM expectations.

Using this infrastructure, limits are placed in three BSM scenarios: 
a simplified model of pair production of weakly interacting dark-matter candidates,
a model with an invisibly decaying Higgs boson, and an effective field theory
with general interactions of electroweak bosons with a dark-matter candidate.
Limits in simplified models are competitive with previous approaches and
the use of shape information in the differential spectra measured in this paper provides 
improved sensitivity to models where the dark-matter candidate mass is close to half the mediator mass.
For the specific effective field theory operators considered in the interpretation, the dark-matter
interactions would evade direct-detection experiments. 
The results presented here represent the most stringent constraints to date on such interactions, with an order-of-magnitude improvement over previous limits from indirect-detection experiments.

The detector-corrected data are published along with the statistical
and systematic uncertainty correlations so that they can easily be used in the future 
to place limits in a wide range of new-physics models that predict final states with jets and missing transverse momentum.

\section*{Acknowledgements}

We thank CERN for the very successful operation of the LHC, as well as the
support staff from our institutions without whom ATLAS could not be
operated efficiently.

We acknowledge the support of ANPCyT, Argentina; YerPhI, Armenia; ARC, Australia; BMWFW and FWF, Austria; ANAS, Azerbaijan; SSTC, Belarus; CNPq and FAPESP, Brazil; NSERC, NRC and CFI, Canada; CERN; CONICYT, Chile; CAS, MOST and NSFC, China; COLCIENCIAS, Colombia; MSMT CR, MPO CR and VSC CR, Czech Republic; DNRF and DNSRC, Denmark; IN2P3-CNRS, CEA-DSM/IRFU, France; SRNSF, Georgia; BMBF, HGF, and MPG, Germany; GSRT, Greece; RGC, Hong Kong SAR, China; ISF, I-CORE and Benoziyo Center, Israel; INFN, Italy; MEXT and JSPS, Japan; CNRST, Morocco; NWO, Netherlands; RCN, Norway; MNiSW and NCN, Poland; FCT, Portugal; MNE/IFA, Romania; MES of Russia and NRC KI, Russian Federation; JINR; MESTD, Serbia; MSSR, Slovakia; ARRS and MIZ\v{S}, Slovenia; DST/NRF, South Africa; MINECO, Spain; SRC and Wallenberg Foundation, Sweden; SERI, SNSF and Cantons of Bern and Geneva, Switzerland; MOST, Taiwan; TAEK, Turkey; STFC, United Kingdom; DOE and NSF, United States of America. In addition, individual groups and members have received support from BCKDF, the Canada Council, CANARIE, CRC, Compute Canada, FQRNT, and the Ontario Innovation Trust, Canada; EPLANET, ERC, ERDF, FP7, Horizon 2020 and Marie Sk{\l}odowska-Curie Actions, European Union; Investissements d'Avenir Labex and Idex, ANR, R{\'e}gion Auvergne and Fondation Partager le Savoir, France; DFG and AvH Foundation, Germany; Herakleitos, Thales and Aristeia programmes co-financed by EU-ESF and the Greek NSRF; BSF, GIF and Minerva, Israel; BRF, Norway; CERCA Programme Generalitat de Catalunya, Generalitat Valenciana, Spain; the Royal Society and Leverhulme Trust, United Kingdom.

The crucial computing support from all WLCG partners is acknowledged gratefully, in particular from CERN, the ATLAS Tier-1 facilities at TRIUMF (Canada), NDGF (Denmark, Norway, Sweden), CC-IN2P3 (France), KIT/GridKA (Germany), INFN-CNAF (Italy), NL-T1 (Netherlands), PIC (Spain), ASGC (Taiwan), RAL (UK) and BNL (USA), the Tier-2 facilities worldwide and large non-WLCG resource providers. Major contributors of computing resources are listed in Ref.~\cite{ATL-GEN-PUB-2016-002}.

\printbibliography

\newpage 
\begin{flushleft}
{\Large The ATLAS Collaboration}

\bigskip

M.~Aaboud$^\textrm{\scriptsize 137d}$,
G.~Aad$^\textrm{\scriptsize 88}$,
B.~Abbott$^\textrm{\scriptsize 115}$,
J.~Abdallah$^\textrm{\scriptsize 8}$,
O.~Abdinov$^\textrm{\scriptsize 12}$$^{,*}$,
B.~Abeloos$^\textrm{\scriptsize 119}$,
S.H.~Abidi$^\textrm{\scriptsize 161}$,
O.S.~AbouZeid$^\textrm{\scriptsize 139}$,
N.L.~Abraham$^\textrm{\scriptsize 151}$,
H.~Abramowicz$^\textrm{\scriptsize 155}$
% ...
\bigskip
\\
% ...
$^{8}$ Department of Physics, The University of Texas at Arlington, Arlington TX, United States of America\\
% ...
$^{12}$ Institute of Physics, Azerbaijan Academy of Sciences, Baku, Azerbaijan\\
% ...
$^{88}$ CPPM, Aix-Marseille Universit{\'e} and CNRS/IN2P3, Marseille, France\\
% ...
$^{115}$ Homer L. Dodge Department of Physics and Astronomy, University of Oklahoma, Norman OK, United States of America\\
% ...
$^{119}$ LAL, Univ. Paris-Sud, CNRS/IN2P3, Universit{\'e} Paris-Saclay, Orsay, France\\
% ...
$^{137}$ $^{(a)}$ Facult{\'e} des Sciences Ain Chock, R{\'e}seau Universitaire de Physique des Hautes Energies - Universit{\'e} Hassan II, Casablanca; $^{(b)}$ Centre National de l'Energie des Sciences Techniques Nucleaires, Rabat; $^{(c)}$ Facult{\'e} des Sciences Semlalia, Universit{\'e} Cadi Ayyad, LPHEA-Marrakech; $^{(d)}$ Facult{\'e} des Sciences, Universit{\'e} Mohamed Premier and LPTPM, Oujda; $^{(e)}$ Facult{\'e} des sciences, Universit{\'e} Mohammed V, Rabat, Morocco\\
% ...
$^{139}$ Santa Cruz Institute for Particle Physics, University of California Santa Cruz, Santa Cruz CA, United States of America\\
% ...
$^{151}$ Department of Physics and Astronomy, University of Sussex, Brighton, United Kingdom\\
% ...
$^{155}$ Raymond and Beverly Sackler School of Physics and Astronomy, Tel Aviv University, Tel Aviv, Israel\\
% ...
$^{161}$ Department of Physics, University of Toronto, Toronto ON, Canada\\
% ...
$^{a}$ Also at Department of Physics, King's College London, London, United Kingdom\\
$^{b}$ Also at Institute of Physics, Azerbaijan Academy of Sciences, Baku, Azerbaijan\\
$^{c}$ Also at Novosibirsk State University, Novosibirsk, Russia\\
$^{d}$ Also at TRIUMF, Vancouver BC, Canada\\
$^{e}$ Also at Department of Physics {\&} Astronomy, University of Louisville, Louisville, KY, United States of America\\
% ...
$^{*}$ Deceased
\end{flushleft}


\end{document}